\documentclass[twocolumn]{aastex631}

\usepackage{amsmath}
\usepackage{booktabs}
\usepackage{enumerate}
\usepackage{tablefootnote}
\usepackage{makecell}
\usepackage{hyperref}

\begin{document}
\title{\texttt{PhDLspec}: physical-prior embedded deep learning method for spectroscopic determination of stellar labels in high-dimensional parameter space}

\author[0000-0002-9332-2871]{Tianmin Wu}
\affiliation{CAS Key Laboratory of Optical Astronomy, National Astronomical Observatories, Chinese Academy of Sciences, Beijing 100101, People’s Republic of China}
\affiliation{School of Astronomy and Space Science, University of Chinese Academy of Sciences, Beijing 100049, People’s Republic of China}

\author[0000-0002-5818-8769]{Maosheng Xiang}\email{msxiang@nao.cas.cn}
\affiliation{CAS Key Laboratory of Optical Astronomy, National Astronomical Observatories, Chinese Academy of Sciences, Beijing 100101, People’s Republic of China}
\affiliation{Institute for Frontiers in Astronomy and Astrophysics, Beijing Normal University, Beijing 102206, People’s Republic of China}

\author{Jianrong Shi}
\affiliation{CAS Key Laboratory of Optical Astronomy, National Astronomical Observatories, Chinese Academy of Sciences, Beijing 100101, People’s Republic of China}
\affiliation{School of Astronomy and Space Science, University of Chinese Academy of Sciences, Beijing 100049, People’s Republic of China}

\author{Meng Zhang}
\affiliation{CAS Key Laboratory of Optical Astronomy, National Astronomical Observatories, Chinese Academy of Sciences, Beijing 100101, People’s Republic of China}
\affiliation{Max-Planck-Institut f\"ur Astronomie, K\"onigstuhl 17, D-69117 Heidelberg, Germany}

\author{Lanya Mou}
\affiliation{CAS Key Laboratory of Optical Astronomy, National Astronomical Observatories, Chinese Academy of Sciences, Beijing 100101, People’s Republic of China}
\affiliation{School of Astronomy and Space Science, University of Chinese Academy of Sciences, Beijing 100049, People’s Republic of China}

\author[0000-0002-8609-3599]{Hong-Liang Yan}
\affiliation{CAS Key Laboratory of Optical Astronomy, National Astronomical Observatories, Chinese Academy of Sciences, Beijing 100101, People’s Republic of China}
\affiliation{Institute for Frontiers in Astronomy and Astrophysics, Beijing Normal University, Beijing 102206, People’s Republic of People’s Republic of China}
\affiliation{School of Astronomy and Space Science, University of Chinese Academy of Sciences, Beijing 100049, People’s Republic of China}

\author{A-Li Luo}
\affiliation{CAS Key Laboratory of Optical Astronomy, National Astronomical Observatories, Chinese Academy of Sciences, Beijing 100101, People’s Republic of China}
\affiliation{School of Astronomy and Space Science, University of Chinese Academy of Sciences, Beijing 100049, People’s Republic of China}




\begin{abstract}
Unlocking the full physical information encoded in low-resolution spectra poses a significant challenge for astronomical survey analysis. Such a task demands modeling spectra and optimizing astrophysical parameters in high-dimensional space, as a consequence of line blending. Here we present \texttt{PhDLspec} -- a deep learning framework embedded with physical priors for stellar spectra modeling and analysis. By imposing differential spectra derived from \textit{ab initio} stellar atmospheric model calculation on a {\sc transformer} framework, \texttt{PhDLspec} can rigorously and precisely model stellar spectra by simultaneously taking into account more than 30 physical parameters, at a computational speed hundreds of times faster than \textit{ab initio} model calculation. With such a flexible stellar modeling approach, \texttt{PhDLspec} can effectively derive $\sim$30 stellar labels from a low-resolution spectrum using affordable optimization techniques. Application to LAMOST spectra ($R\lesssim1800$) yields stellar elemental abundances in good agreement with high-resolution spectroscopic surveys, following essential calibrations to correct systematic biases in elemental abundance estimates using wide binaries and reference high-resolution datasets. We provide a catalog of 25 elemental abundances for 116,611 subgiant stars with precise age estimates. The successful application of \texttt{PhDLspec} to LAMOST spectra for high-dimensional parameter determination sheds light on similar challenges faced by other surveys and disciplines.
\end{abstract}


\section{Introduction} \label{sec:intro}

As artificial intelligence models become increasingly prevalent for inferring astrophysical parameters from observational datasets generated by astronomical surveys, the physical interpretability of the models has emerged as a fundamental challenge in data analysis. A robust and physics-grounded determination of astrophysical parameters generally involves two critical components: first, the construction of a transparent and physics-based forward model that both accurately represents the observed data and is differentiable to the model parameters; and second, the application of a sophisticated parameter inference framework capable of reliably estimating uncertainties and covariances among the derived parameters.

These are challenging tasks for stellar parameter determination from low-resolution spectra, as they require spectral modeling and parameter inference in a high-dimensional parameter space -- usually with a dimension higher than 20 -- as a consequence of line blending \citep{2017ApJ...849L...9T}. 
Limited by computational costs, such as the affordable size of spectral templates as well as the speed and precision of model spectrum interpolation, it was a common and practical way in the pipeline of determining parameter for low-resolution survey spectra to consider only a few basic stellar parameters, including effective temperature $T_{\rm eff}$, surface gravity $\log g$, overall metallicity [M/H] ([Fe/H] is used in many cases), and, perhaps, the $\alpha$-to-iron abundance ratio [$\alpha$/Fe] \citep[e.g.][]{2011AJ....141...90L, 2014IAUS..306..340W, 2015MNRAS.448..822X, 2016RAA....16..110L, 2023ApJ...947...37C}, while leaving individual elemental abundances undetermined.   

Various approaches have been proposed to derive more stellar labels from low-resolution and blended spectra from large surveys. One way is to perform label transfer by mapping a low-resolution spectrum to one or more labels from high-resolution spectroscopy with a machine learning approach \citep[e.g.][]{2017MNRAS.464.3657X, 2020ApJ...891...23W}. Such methods can predict stellar labels for a large number of stars in a short time, but they usually operate as a black box that lacks transparency and physical interpretability. As a result, the derived parameters are likely estimated via astrophysical correlations rather than physics-grounded measurements and determinations. 

Another way is to build a forward model between stellar labels and stellar spectra, and determine the labels simultaneously through spectral fitting \citep[e.g.][]{2009A&A...501.1269K, 2015ApJ...808...16N, 2016ApJ...826L..25R}. The development of this forward modeling approach has particularly benefited from the usage of machine learning and deep learning methods, which have demonstrated high flexibility in building precise spectral models in high-dimensional parameter space \citep[e.g.][]{2019ApJ...879...69T, 2022A&A...662A..66X, 2025ApJS..279....5Z, 2025PASA...42...51B}.


It has been demonstrated that machine learning and deep learning methods in a data-driven framework have great power for stellar label determination \citep[e.g.][]{2015ApJ...808...16N, 2020ApJ...898...58W}, primarily owing to their ability to provide self-consistent models that effectively represent the survey spectra. However, due to complex astrophysical correlations among stellar labels, as well as measurement errors in spectral flux, data-driven models may fail to capture the true (physical) features of a given parameter. It is therefore necessary to regularize the training process of a data-driven model with physical priors of spectral features to ensure physics-grounded stellar parameter determination \citep{2017ApJ...849L...9T, 2019ApJS..245...34X}. Such a hybrid forward modeling approach, combining data-driven and physical priors, has enabled physics-grounded determination of multiple stellar abundances for millions of stars from low-resolution survey spectra \citep{2019ApJS..245...34X, 2024ApJS..273...19Z, 2025ApJS..279....5Z}. 
Nonetheless, data-driven approaches have their own limitations. One drawback lies in their dependence on — and thus confinement to — a training set with known labels, which often imperfectly represent the full survey data. For instance, when using stellar labels from high-resolution spectroscopic surveys such as APOGEE \citep{2017AJ....154...94M, 2022ApJS..259...35A} and GALAH \citep{2015MNRAS.449.2604D, 2021MNRAS.506..150B} as training sets to estimate labels from the massive low-resolution spectra of LAMOST \citep{2012RAA....12.1197C, 2022Innov...300224Y}, only FGK-type stars can be reliably handled, while regimes of low metallicity or high effective temperature are poorly constrained \citep[e.g.][]{2019ApJS..245...34X, 2025ApJS..279....5Z}. In addition, stellar labels from high-resolution surveys themselves may contain systematic errors, which are inevitably propagated into low-resolution spectral analysis under a data-driven framework \citep[e.g.][]{2019ApJS..245...34X}.

To alleviate the reliance on high-quality empirical training sets with known stellar labels, one may adopt either a model-driven approach, which uses synthetic spectra to construct the spectral model \citep[e.g.,][]{2019ApJ...879...69T, 2022A&A...662A..66X, 2025ApJ...980...66R}, or a hybrid approach that maps synthetic spectra onto the observed domain via domain adaptation techniques \citep{2021ApJ...906..130O, 2023ApJS..266...40W, 2025ApJS..277...47Z}. The latter is specifically designed to mitigate the impact of systematic errors inherent in synthetic spectra. However, the effectiveness of these frameworks in the context of low-resolution spectroscopy has so far been demonstrated only for the recovery of a small number of labels, whereas spectral modeling and parameter inference in high-dimensional spaces remain substantially more challenging.


Here we introduce \texttt{PhDLspec}, a physics-grounded deep learning framework designed for high-dimensional stellar spectral modeling and parameter estimation. Based on the {\sc transformer} architecture \citep{2017arXiv170603762V}, \texttt{PhDLspec} accurately emulates \textit{ab initio} spectra from the \texttt{Kurucz} ATLAS model \citep{1970SAOSR.309.....K, 1993sssp.book.....K, 2005MSAIS...8...14K} across more than 30 parameters, while operating hundreds of times faster than \textit{ab initio} spectral synthesis. A key ingredient of \texttt{PhDLspec} is the incorporation of \textit{ab initio} gradient spectra, defined as the first-order derivatives of the flux with respect to stellar labels, as physical priors to regularize the training loss, ensuring that the model reproduces both flux values and their first-order derivatives. This is essential for the model to make interpretable and physically self-consistent inference. With such a flexible and physics-grounded spectral model, \texttt{PhDLspec} employs the CMA-ES algorithm \citep{6790628} to efficiently fit observed spectra to estimate the stellar labels, achieving an optimal trade-off between precision and computational cost.

Applying \texttt{PhDLspec} to LAMOST low-resolution spectra ($R \lesssim 1800$) reliably infers 28 stellar labels, including atmospheric parameters ($T_{\rm{eff}}$, $\log g$, [Fe/H]), light elements (C, N, Na, Al), $\alpha$-elements (O, Mg, Si, Ca, Ti), iron-peak elements (Sc, V, Cr, Mn, Co, Ni), and neutron-capture elements (Sr, Y, Zr, Ba, La, Ce, Pr, Nd, Sm, Eu). While the abundances show broad consistency with high-resolution surveys, residual systematic biases remain. These are mitigated through a tailored flagging system and automated calibration using wide binaries and reference high-resolution datasets.

The results highlight the potential of low-resolution spectra for precise chemical abundance determination. Validation with wide binaries and open cluster member stars indicates that the abundance precision reaches 0.1~dex for $\alpha$-elements and iron-peak elements, improving to 0.05~dex for a number of them, and 0.2-0.3~dex for neutron-capture elements. We provide a catalog of 25 elemental abundances derived from LAMOST spectra for 116,611 subgiant stars with precise age estimates. The catalog constitutes a valuable new dataset for the studies of Galactic archaeology.

The paper is organized as follows. Section \ref{sec:method} details the architecture and workflow of the \texttt{PhDLspec} model. Section \ref{sec:modeltraining} presents the model training procedure and evaluates its performance. Calibration and validation of the \texttt{PhDLspec} method are carried out in Section \ref{sec:calibration} using wide binaries, the M67 open cluster, and high-resolution spectroscopic data. Section \ref{sec:subgiant catalog} releases a value-added catalog of elemental abundances from LAMOST low-resolution spectra for 116,611 subgiant stars with precise age estimates. The paper concludes with a summary in Section \ref{sec:summary}.

\begin{figure*}[htbp]
		\centering
		\includegraphics[width=0.9\linewidth]{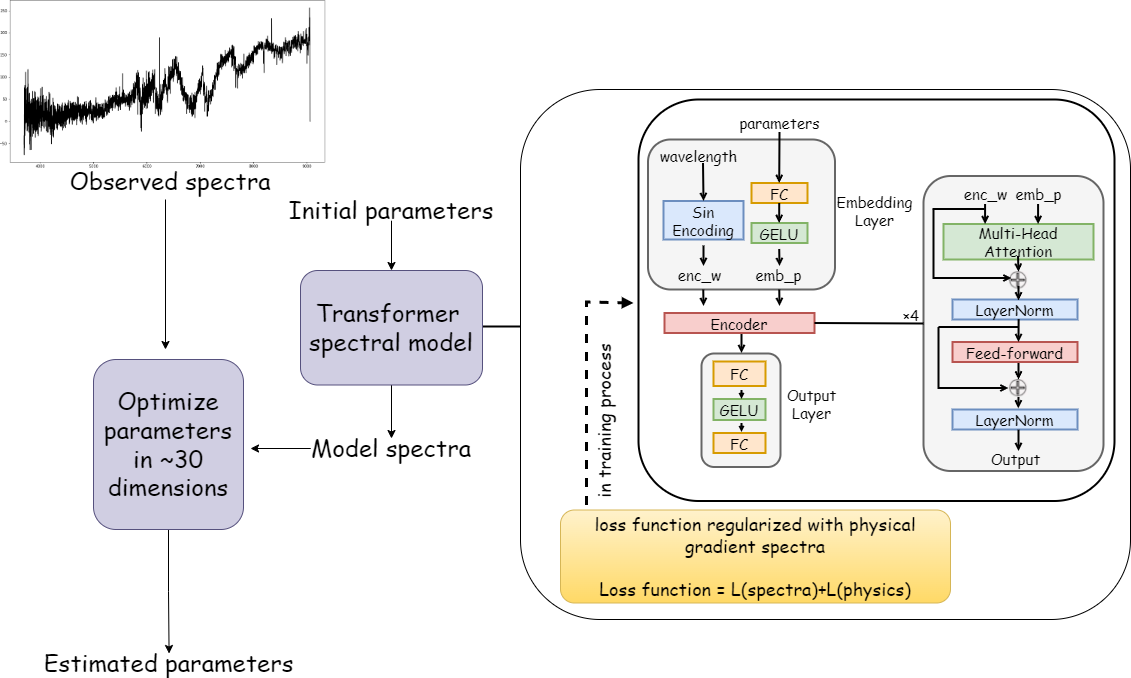}
		\caption{Schematic diagram of the \texttt{PhDLspec} method. The transformer-based spectral model employs an embedding layer that combines wavelength encoding and parameter embedding, followed by multiple encoder blocks with multi-head attention, layer normalization, and feed-forward layers. The output spectra are generated through a fully connected output layer. The model training is regularized with physical gradient spectra from \textit {ab initio} computation. Observed spectra are fitted by simultaneously optimizing stellar parameters in a $\sim$30-dimensional parameter space.
}
       \label{fig:flow}
\end{figure*}

\section{\texttt{PhDLspec}} \label{sec:method}

\texttt{PhDLspec} performs spectroscopic analysis and determines stellar parameters using a forward modeling approach. It consists of two key components: first, a flexible and physics-grounded spectral modeling algorithm capable of accurately interpolating spectra across over 30 dimensions, represented as $\mathbf{\theta} \rightarrow f(\lambda; \mathbf{\theta})$, achieved with a Transformer-based framework \citep{2017arXiv170603762V}. 
To ensure the model faithfully captures the physical behavior of each parameter, gradient spectra $\nabla_\theta{f(\lambda; \mathbf{\theta})}$ -- defined as the partial derivatives of the flux with respect to the parameters $\frac{\partial f(\lambda;\theta)}{\partial \theta}$ -- computed with the \texttt{Kurucz} ATLAS12 atmospheric model \citep{2005MSAIS...8...25C} are incorporated as physical constraints during training.

The second component of \texttt{PhDLspec} is a high-dimensional parameter fitting scheme that enables simultaneous estimation of up to 30 parameters from an observed spectrum: $f_{\text{obs}}(\lambda) \rightarrow \hat{\mathbf{\theta}}$. This is implemented using the CMA-ES algorithm \citep{2016arXiv160400772H}. A flowchart illustrating the overall structure of \texttt{PhDLspec} is provided in Figure~\ref{fig:flow}.


\subsection{Spectral modeling}
\label{subsec:spectral model}
\subsubsection{Transformer-based encoder model}
The {\sc transformer} model \citep{2017arXiv170603762V} has been widely applied to the area of natural language processing (NLP). It revolutionizes sequence modeling by eliminating the need for recurrent or convolutional layers and relying entirely on self-attention mechanisms to capture dependencies between elements in a sequence. Thus, it allows for efficient parallel processing and long-range contextual understanding. The core component of the {\sc transformer} is the multi-head self-attention mechanism, which enables the model to focus on different parts of the sequence simultaneously, enhancing its ability to model complex relationships in data. 

In astronomical studies, {\sc transformer} has been shown to be superior to simple neural networks in processing long-range information from spectra \citep{2024arXiv240705751R}. Its attention mechanism is particularly well-suited for capturing the spectral lines of various chemical elements blended in low-resolution spectra. In this work, we adopt a transformer-based encoder model to learn the intricate relationship between stellar parameters and spectra. As illustrated in the Transformer spectral model block in Figure~\ref{fig:flow}, the model comprises three main components: an embedding layer, four encoder blocks, and an output layer. The detailed architecture of the model is described below.

\begin{enumerate}[(1)]
\item Embedding layer:
The data fed into the encoder blocks typically needs to be transformed into a format that the computer can efficiently process, which is achieved through input embedding. In NLP, one-hot encoding and word embedding are often used to convert words into vectors. For our spectral model, we employ a fully connected layer with 128 neurons followed by a Gaussian Error Linear Unit (GELU, \citet{hendrycks2023gaussianerrorlinearunits}, Equation~\ref{eq:gelu}) activation function to embed the parameters ($\theta$) that serves as an input to the encoder. Compared with the Rectified Linear Unit (ReLU, \citet{Glorot2011DeepSR}), GELU is smoother and differentiable everywhere, and thus can reduce the risk of gradient explosions.
\begin{equation}
\begin{aligned}
\text{GELU}(x)=0.5x(1+\tanh{(\sqrt{\frac{2}{\pi}}(x+0.044715x^3)))}.
\label{eq:gelu}
\end{aligned}
\end{equation}
In addition, since the {\sc transformer} is computed in parallel, positional encoding is necessary to record the positional information of each word. Here we represent the wavelength ($\lambda$) by sinusoidal encoding (Equation~\ref{eq:sin encoding}), feeding it as another input to the encoder. This allows the model to interpret spectra sampled on different wavelength grids within a single training framework \citep{2024arXiv240705751R}.
\begin{equation}
\begin{aligned}
\lambda_{\rm enc}=\mathrm{sin}(\frac{2\pi\lambda}{T_d}),
\label{eq:sin encoding}
\end{aligned}
\end{equation}
where $T_{d}$ denotes different cycles, which are obtained by sampling uniformly from $T_{\mathrm{min}}=1e-6$ to $T_{\mathrm{max}}=10$ in logarithmic space also with d=128 dimensions (Equation \ref{eq:cycle}).
\begin{equation}
\begin{aligned}
T_{d}=\text{logspace}[\mathrm{lg}(T_{\rm min}),\mathrm{lg}(T_{\rm max}), \text{steps}=d]
\label{eq:cycle}
\end{aligned}
\end{equation}
\item Encoder blocks:
The encoded wavelength and parameters are then passed through several stacked encoder blocks. Each encoder contains a multi-head attention mechanism and a feed-forward layer, and performs residual connection and layer normalization (Add\&Norm) after these two layers. 
The attention mechanism serves as the core component of the encoder. Compared to self-attention, the multi-head attention mechanism enables each head to operate in distinct subspaces, learning independent representations of the input. In each head, we compute the query (Q) by linearly transforming the wavelength, while the key (K) and value (V) are obtained through the linear transformation of the parameters. 
\begin{equation}
\begin{aligned}
Q&=\lambda_{\mathrm{enc}}W^Q,\\
K&=p_{\mathrm{emb}}W^K,\\
V&=p_{\mathrm{emb}}W^V,
\label{eq:qkv}
\end{aligned}
\end{equation}
where $W^Q$,$W^K$,$W^V$is the weight matrices.
The attention of Q, K, V of the $i$th head is computed by Equation \ref{eq:attention}:
\begin{equation}
\begin{aligned}
\text{Attention}(Q_{i},K_{i},V_{i})=\mathrm{softmax}(\frac{Q_{i}K_{i}^T}{\sqrt{d_{k}}})V_{i},
\label{eq:attention}
\end{aligned}
\end{equation}
where $d_{k}$ is the feature dimension.
The outputs of all heads are connected and mapped through a linear layer to get the final result (Equation \ref{eq:multiatten}). 
\begin{align}
\text{MultiHead}(Q,K,V)&=\text{concat}(A_{1},...,A_{h})W^O,\\
A_{i}&=\text{Attention}(Q_{i},K_{i},V_{i}),
\label{eq:multiatten}
\end{align}
where $W^O$ is the output mapping matrix and $h$ is the number of heads.
By combining multiple attention heads, the model can obtain richer and more complex representations, thereby enhancing its ability to capture the intricate relationships between the parameters and the spectra. In our work, we choose to use four encoder blocks and four heads for the multi-head attention mechanism.

The residual connection follows the concept of the Residual Neural Network \citep{he2015deepresiduallearningimage}, where the input $w_{\mathrm{enc}}$ to the Multi-Head Attention is directly added to the output, allowing the network to learn deeper feature representations. The result is then normalized using Layer Normalization \citep{ba2016layernormalization}. The Add\&Norm operation helps mitigate the vanishing gradient problem, making training more stable and efficient. Following this, the data pass through a nonlinear transformation via a feed-forward neural network and a subsequent Add\&Norm step, yielding the final output matrix.

\item Output layer:
The output layer consists of two fully connected layers with 256 neurons, with the GELU activation function applied after the first layer.

\end{enumerate}

\subsubsection{Regularize the model with differential spectra}
Given a set of stellar parameters $\mathbf{\theta}$, a physics-grounded spectral model should be able to accurately generate not only the spectral flux $f(\lambda; \mathbf{\theta})$, but also the ground-truth flux gradients $\nabla_\theta{f(\lambda; \mathbf{\theta})}$, the latter representing the physical features of the parameters. Due to the degeneracy among the features of different parameters, this is not an obvious or directly achievable task, but rather requires dedicated effort.  

To achieve this goal, we adopt the idea of \texttt{DD-Payne} that uses the {\textit {ab initio}} gradient spectra as a prior to regularize the model training process, as initially proposed by \citet{2017ApJ...849L...9T}. This is implemented by defining the loss function $L$ as 
\begin{equation}
L = L_{\mathrm {spec}} + L_{\mathrm {grad}},
    \label{loss}
\end{equation}

\begin{equation}
\begin{aligned}
L_{\mathrm {spec}}:=\frac{1}{N}\sum_{i=0}^{N-1}|f_{i}(\lambda)-f_{i}(\lambda;\theta)|,
\label{specloss}
\end{aligned}
\end{equation}

\begin{equation}
\begin{aligned}
L_{\mathrm {grad}}:=\frac{1}{N}\sum_{i=0}^{N-1}\sum_{m=0}^{M-1}\omega_i\times\vert\nabla_{\theta_m}{f_i(\lambda)}-\nabla_{\theta_m}f_i(\lambda; \mathbf{\theta}_m)\vert,
\label{gradspecloss}
\end{aligned}
\end{equation}

where $N$ is the number of samples contained in the dataset, and $M$ is the number of labels to be determined. $f(\lambda)$ denotes the {\textit{ab initio}} model spectrum, and $f(\lambda;\theta)$ is the spectrum predicted by \texttt{PhDLspec}. To prevent elements with intrinsically larger gradient amplitudes from dominating the optimization, we introduce a per-element weight defined as $w_{i}=\frac{0.1}{\sigma_{i}}$, where $\sigma_{i}=\mathrm{Std}_{\lambda}(\nabla_{\theta_m}{f_i(\lambda)})$ denotes the standard deviation of the elemental gradient spectrum along the wavelength dimension. 
This weighting effectively compensates for the fact that some elements exhibit stronger spectral responses than others, ensuring that elements with weak spectral features contribute comparably to the loss function as those with strong features. Note that, unlike \citet{2019ApJS..245...34X} and \citet{2024arXiv240705751R} who adopt the L2 loss, we choose the L1 loss simply because $L_{\mathrm {spec}}$ and $L_{\mathrm{grad}}$ can share the same physical meaning and are therefore directly comparable. In addition, the L1 loss is less sensitive to outliers.


While the power of such a regularization strategy has been demonstrated in data-driven models \citep{2017ApJ...849L...9T, 2019ApJS..245...34X}, in this study we show that even in a model-driven deep learning approach, such a regularization strategy is essential for building a physics-grounded model in high-dimensional parameter space. The model is trained by minimizing the total loss defined above. Detailed descriptions of the model training, such as the optimizer, learning rate, and stopping criteria are presented in Section~\ref{subsec:trainandval}. 


\subsection{Parameter estimation with spectra fitting}

\texttt{PhDLspec} estimates stellar parameters by fitting observed spectra to models. This fitting process, when employing the aforementioned transformer-based spectral model, presents a high-dimensional, non-convex optimization problem. Moreover, the function's analytical form is unknown, making it what is mathematically known as a stochastic black-box optimization problem. While established methods such as Markov Chain Monte Carlo (MCMC), Bayesian optimization, and genetic algorithms can eventually converge to an optimum, their efficiency is severely limited by computational demands. To address this, we employ the Covariance Matrix Adaptation Evolution Strategy (CMA-ES), an efficient evolutionary algorithm particularly well-suited for such complex challenges \citep{6790628}.

\subsubsection{The CMA-ES method}

CMA-ES is a global optimization method based on evolutionary algorithms that adapts to the properties of the objective function by dynamically adjusting the covariance matrix of the search distribution \citep{2016arXiv160400772H}. This allows it to operate efficiently in complex search spaces. 

Similar to most evolutionary algorithms, CMA-ES includes several processes: population initialization, mutation, evolution, selection, and reorganization. The specific procedures are detailed below:
\begin{enumerate}[(1)]
\item Initialization: The initial population is first generated by a multivariate normal distribution $\mathcal N(m_{0},\sigma^2_{0}I)$. $m_{0}$ is the mean value of the initial population, $\sigma_{0}$ is the global variance or step size, determining the search area. 
\item Mutation: The mutation process of CMA-ES is centered on the mean value of the current generation, which is computed as:
\begin{equation}
\begin{aligned}
x_{k}^{(g+1)}\sim m^{(g)}+\sigma^{(g)}\mathcal N(0,C^{(g)})\:\mathrm{for}\,k=1,...,\lambda
\end{aligned}
\label{eq:cma}
\end{equation}
where $x_{k}^{(g+1)}$ is the $k$th individual in the $g$+1-th generation, $m^{(g)}$ and $\sigma^{(g)}$ is the mean value and the step size of the $g$-th distribution. 
\item Evolution and Selection: Select $\mu$ individuals with the optimal fitness function from the $\lambda$ individuals of the $g$-th generation. These $ \mu$ individuals will serve as the parents of the next generation distribution.
\item Reorganization: Use the information from the optimal individuals of the $g$-th generation to update the algorithm's strategy parameters, including the step size $\sigma$, mean value $m$, and covariance matrix $C$. Then, generate individuals in the next generation distribution using mutation operations based on these parameters. The updating expression of $\sigma$, $m$, and $C$ is as follows:
\begin{equation}
\begin{aligned}
\sigma^{(g+1)}=\sigma^{(g)}\exp({\frac{c_{\sigma}}{d_{\sigma}}(\frac{\|p_{\sigma}^{(g+1)}\|}{E\|\mathcal N(0,I)\|}-1))}
\end{aligned}
\end{equation}

\begin{equation}
\begin{aligned}
m^{(g+1)}=\sum_{i=1}^{\mu}w_{i}x^{(g+1)}_{i:\lambda}
\end{aligned}
\end{equation}

\begin{equation}
\begin{aligned}
C^{(g+1)}=(1-c_{\mu}\sum w_{i})C^{(g)}+c_{\mu}\sum_{i=1}^{\lambda}w_{i}y_{i:\lambda}^{(g+1)}y_{i:\lambda}^{(g+1)^T}
\end{aligned}
\end{equation}
\end{enumerate}

The fitting procedure is formulated as the minimization of a reduced $\chi^2$ objective function that quantifies the difference between the observed and model spectra:

\begin{equation}
\begin{aligned}
\mathcal{X}^2(\theta)=\frac{1}{n}\sum_{i=1}^n \left(\frac{f_{\mathrm{obs},i}(\lambda)-f_i(\lambda;\theta)}{\sigma_i(\lambda)}\right)^2,
\end{aligned}
\label{eq:objectivefunction}
\end{equation}
where $f_{\mathrm{obs}}(\lambda)$ is the observed spectrum, $f(\lambda;\theta)$ is the model prediction, $\sigma(\lambda)$ is the observational uncertainty, and $n$ is the number of pixels. The optimization is carried out using CMA-ES, implemented with the Python package \texttt{pycma} \citep{hansen2019pycma}.

Although \texttt{PhDLspec} is designed for simultaneous optimization of all labels during spectral fitting, a multi-stage strategy can be adopted to reduce dimensionality and accelerate the process when computational resources are limited. In this work, we apply a two-stage fitting approach to LAMOST spectra. 
In the first stage, a subset of parameters with strong spectral features is optimized while the remaining parameters are kept at their initial values. In the second stage, the remaining parameters are fitted based on the results obtained in the first stage.
For the LAMOST spectra, parameters are divided into two groups according to the strengths of their spectral sensitivities:
\begin{description}
  \item[Group I] $T_{\rm eff}$, $\log g$, [Fe/H], [C/Fe], [N/Fe], [O/Fe], [Na/Fe], [Mg/Fe], [Al/Fe], [Ca/Fe], [Ti/Fe], [V/Fe], [Cr/Fe], [Mn/Fe], [Co/Fe], [Ni/Fe];
  \item[Group II] [Si/Fe], [Sc/Fe], [Sr/Fe], [Y/Fe], [Zr/Fe], [Ba/Fe], [La/Fe], [Ce/Fe], [Pr/Fe], [Nd/Fe], [Sm/Fe], [Eu/Fe].
\end{description}

The optimization is constrained by predefined parameter bounds, $[-0.25,1.25]$. To ensure that sampled parameters remain within this range, we employ the \texttt{BoundTransform} function in \texttt{pycma}, which maps candidate solutions back to the feasible region through a smooth transformation (e.g., sigmoid). 

\subsubsection{Parameter uncertainty estimation}

The uncertainty of parameters fitted by \texttt{pycma} is given by $\sigma \cdot \sqrt{C_{ii}}$, where $\sigma$ is the step size and $C$ is the covariance matrix at the final iteration. While this quantity reflects the scale of the exploration in the final phase of the CMA-ES search, it does not strictly represent the true posterior covariance of the parameters. Moreover, in cases of strong convergence, the estimated uncertainties can become exceedingly small - often on the order of $10^{-5}$ - which may be unrealistically precise.

Therefore, to obtain more reliable uncertainties of the parameters $\sigma(\theta)$, we manually compute the covariance matrix of the high-dimensional parameters. This matrix is derived from the Jacobian matrix and the flux errors, as given by Equation \ref{eq:covmatrix}.
\begin{equation}
\begin{aligned}
\sigma(\theta_j)=\sqrt {\text{Cov}(\theta) _{jj}},
\end{aligned}
\end{equation}

\begin{equation}
\begin{aligned}
\text{Cov}(\theta)=(J^{T}WJ)^{-1},
\end{aligned}
\label{eq:covmatrix}
\end{equation}
where $W$ is the diagonal matrix composed of flux errors, and $J_{ij}$ is the transposed matrix of the gradient spectrum. 
\begin{equation}
\begin{aligned}
W=\mathrm{diag}\left(\frac{1}{\sigma_{i}^{2}}\right)
\label{eq:W}
\end{aligned}
\end{equation}

\begin{equation}
\begin{aligned}
J_{ij} = \nabla_{\theta_j}{f(\lambda_i; \mathbf{\theta})} 
\end{aligned}
\end{equation}

\begin{equation}
\begin{aligned}
\nabla_{\theta_j}{f(\lambda_i; \mathbf{\theta})} =\frac{f_i(\lambda;\theta_j+h)-f_i(\lambda;\theta_j-h)}{2h}.
\end{aligned}
\label{eq:1order dev}
\end{equation}

Note that the error estimation discussed in this paper only accounts for the formal error arising from the measurement uncertainties in the spectral flux, and does not include any systematic errors present in either the observed spectra or the \texttt{Kurucz} model spectra. Such systematic errors may cause an imperfect match between the model and the observed spectra, resulting in a $\chi^2$ value greater than unity for the best fit. For an error estimate that better reflects the quality of the spectral fit, one may adopt a modified weight to replace Equation~\ref{eq:W} when computing the error, as follows:
\begin{equation}
\begin{aligned}
W^\prime = \mathrm{diag}\left( \frac{1}{\sigma_i^2 \chi^2} \right).
\end{aligned}
\end{equation}
Both the formal error estimates and the $\chi^2$ value of the best fit are provided by \texttt{PhDLspec}, allowing users to scale the formal error by the $\chi^2$ value if needed.

\subsection{Data flag}
Although the CMA-ES algorithm aims to efficiently approximate the global minimum of the objective function, it may still converge to a local optimum in practical applications. Therefore, we performed an independent minimum $\chi^2$ test for each label estimate. In doing so, 50 sampling points were selected at equal intervals across the normalized abundance range of each element, while the remaining parameters were kept fixed at their CMA-ES best-fit values. The minimum $\chi^2$ value obtained from this scan was taken as the reference optimal solution. 

Based on the above results and CMA-ES results, we introduce a diagnostic indicator represented by a four-digit binary code to flag the derived chemical abundances with reduced reliability. The four components of this flag are described in detail below.

\begin{description}
\item[$F_1$ - CMA-ES boundary flag]
$F_1=0$ if the abundance results are remained within the corresponding parameter space; $F_1=1$ if the solution reaches or exceeds the boundaries of the model grid. Boundary solutions typically indicate reduced reliability and are therefore flagged. 
\item[$F_2$ - $\chi^2$ boundary flag]
$F_2 = 0$ if the minimum $\chi^2$ value from an independent scan falls within the allowed abundance range; $F_2 = 1$ if it reaches the boundary.
\item[$F_3$ - difference $>$ 3$\sigma$]
$F_3 = 0$ if the difference between the CMA-ES abundance and the independent $\chi^2$ estimate is within $3\sigma$, and $F_3 = 1$ if the difference exceeds $3\sigma$. This flag highlights statistically significant discrepancies between the two estimates.
\item[$F_4$ - difference $>$ 0.3 dex]
$F_4 = 0$ if the absolute difference between the CMA-ES and $\chi^2$ abundances is less than 0.3 dex, and
$F_4 = 1$ if the difference is equal to or greater than 0.3 dex. This flag provides an additional threshold to identify potentially unreliable abundances.
\end{description}

In the application to LAMOST data (Section~\ref{sec:calibration} and Section~\ref{sec:subgiant catalog}), high-quality flags ($F_1$–$F_4 = 0$) are obtained in most elements ($\gtrsim90\%$, reaching $\sim99\%$ for Mg, Sc, Ti, Cr, Mn, Co, Ni, Sr, and Ba), intermediate levels for Y, Sm, and Pr ($\sim82$–85\%), and lower fractions for heavy elements such as Zr, Ce, and Eu ($60\%$-$\sim70\%$). Overall, these statistics demonstrate that the majority of derived elemental abundances are robust, with only a small subset of heavy elements (Zr, Ce, and Eu) showing reduced reliability mainly due to their weak spectral lines.

\section{Model training and validation}\label{sec:modeltraining}
\subsection{Training data}\label{kurucz}
We utilize the ATLAS12 code \citep{2005MSAIS...8...25C} to compute the \texttt{Kurucz} 1D LTE stellar atmosphere models, and employ the SYNTHE code \citep{2005MSAIS...8...14K} to solve the radiative transfer equation and generate synthetic spectra for a grid of stars, whose atmospheric parameters are sampled based on the MIST isochrones \citep{2016ApJ...823..102C} within the range of $T_{\rm{eff}}\in$[3800 K, 8000 K] and [Fe/H]$\in$[$-$4~dex, 0.5~dex]. [Fe/H] is uniformly sampled at intervals of 0.1~dex, while $T_{\rm{eff}}$ and $\log g$ are sampled randomly from the isochrones. Values of the other parameters of consideration, including micro-turbulent velocity $v_{\rm mic}$ and abundances for 30 elements, namely, C, N, O, Na, Mg, Al, Si, S, K, Ca, Sc, Ti, V, Cr, Mn, Co, Ni, Cu, Zn, Sr, Y, Zr, Ba, Au, La, Ce, Pr, Nd, Sm, and Eu are uniformly drawn within a pre-defined ranges (Table \ref{tab:table1}).

\begin{table}[htbp]
\centering
  \caption{The pre-defined range of the 30 elements.}
  \label{tab:table1}
\setlength{\tabcolsep}{3mm}
    \begin{tabular}{cc}
\toprule
Elements([X/Fe])            & Range   \\ 
\midrule
C          & [-1.5,1.5] \\
N          & [-1.5,1.5]\\
O& [-0.5,1]\\
Na&[-1,1]\\
Mg& [-0.4,0.6] \\
Al& [-1,1] \\
Si& [-0.4,0.6] \\
S& [-0.4,0.6] \\
K&[-1,1]\\
Ca& [-0.4,0.6]\\
Sc& [-0.8,0.8] \\
Ti& [-0.4,0.6] \\
V& [-0.8,0.8]  \\
Cr & [-0.8,0.8] \\
Mn& [-0.8,0.8] \\
Co& [-0.8,0.8] \\
Ni& [-0.8,0.8] \\
Cu& [-0.8,0.8] \\
Zn& [-0.8,0.8] \\
Sr&[-1.5,2]\\
Y& [-1.5,2] \\
Zr& [-1.5,2]\\
Ba& [-1.5,2]  \\
Au& [-1.5,2]  \\
La& [-1.5,2]\\
Ce&  [-1.5,2] \\
Pr& [-1.5,2]\\
Nd & [-1.5,2]\\
Sm & [-1.5,2]\\
Eu&  [-1.5,2]\\
\bottomrule
\end{tabular}
\end{table}

\begin{figure}[htbp]
		\centering
		\includegraphics[width=1\linewidth]{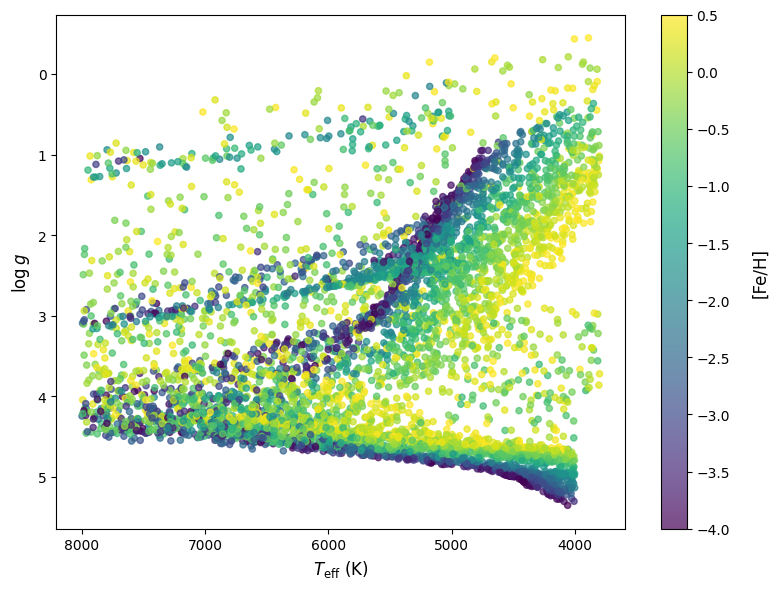}
		\caption{Distribution of the \texttt{Kurucz} spectra training set in the $T_{\rm eff}$-$\log g$ diagram, color-coded by their [Fe/H].}
       \label{fig:parameterspace}
\end{figure}

Ultimately, we obtain a library of 6,222 {\textit {ab initio}} model spectra, self-consistently computed with ATLAS12 and SYNTHE. Figure \ref{fig:parameterspace} shows the distribution of their atmospheric parameters in the $T_{\rm eff} - \log g$ diagram.

For applying \texttt{PhDLspec} to the LAMOST low-resolution spectra, we degrade the resolution of the \texttt{Kurucz} model spectra to match the LAMOST line spread function. The wavelengths are sampled on a uniform grid with a step size of $1\times10^{-4}$ in logarithmic scale, in the range of 3750 -- 8996{\AA}. Similar to previous work \citep{2022A&A...662A..66X}, we normalize the \texttt{Kurucz} spectra by dividing a smoothed version of themselves by convolving a Gaussian kernel, 
\begin{equation}
\begin{aligned}
{f_n}(\lambda)=\frac{f(\lambda)}{\overline{f}(\lambda)},
\end{aligned}
\end{equation}
\begin{equation}
\begin{aligned}
\overline{f}(\lambda_i)=\frac{\sum_{j}\omega_{ij}f(\lambda_j)}{\sum_{j}\omega_{ij}},
\end{aligned}
\label{eq:gaussian}
\end{equation}
\begin{equation}
\begin{aligned}
\omega_{ij} := \text{exp}(-\frac{(\lambda_{j}-\lambda_{i})^2}{L^2}),L=50{\AA}.
\end{aligned}
\label{eq:normkurucz}
\end{equation}

In addition, all of the 34 parameters are standardized according to Equation~\ref{eq:norm} to accelerate model convergence and reduce computational complexity:
\begin{equation}
\begin{aligned}
p^{\prime}=\frac{p-p_{\rm min}}{p_{\rm max}-p_{\rm min}},
\end{aligned}
\label{eq:norm}
\end{equation}
where $p_{\rm min}$ and $p_{\rm max}$ denote the minimum and maximum value of parameters in the training set.

\subsection{Training and validation}\label{subsec:trainandval}
\begin{figure*}[htbp]
		\centering
		\includegraphics[width=0.95\linewidth]{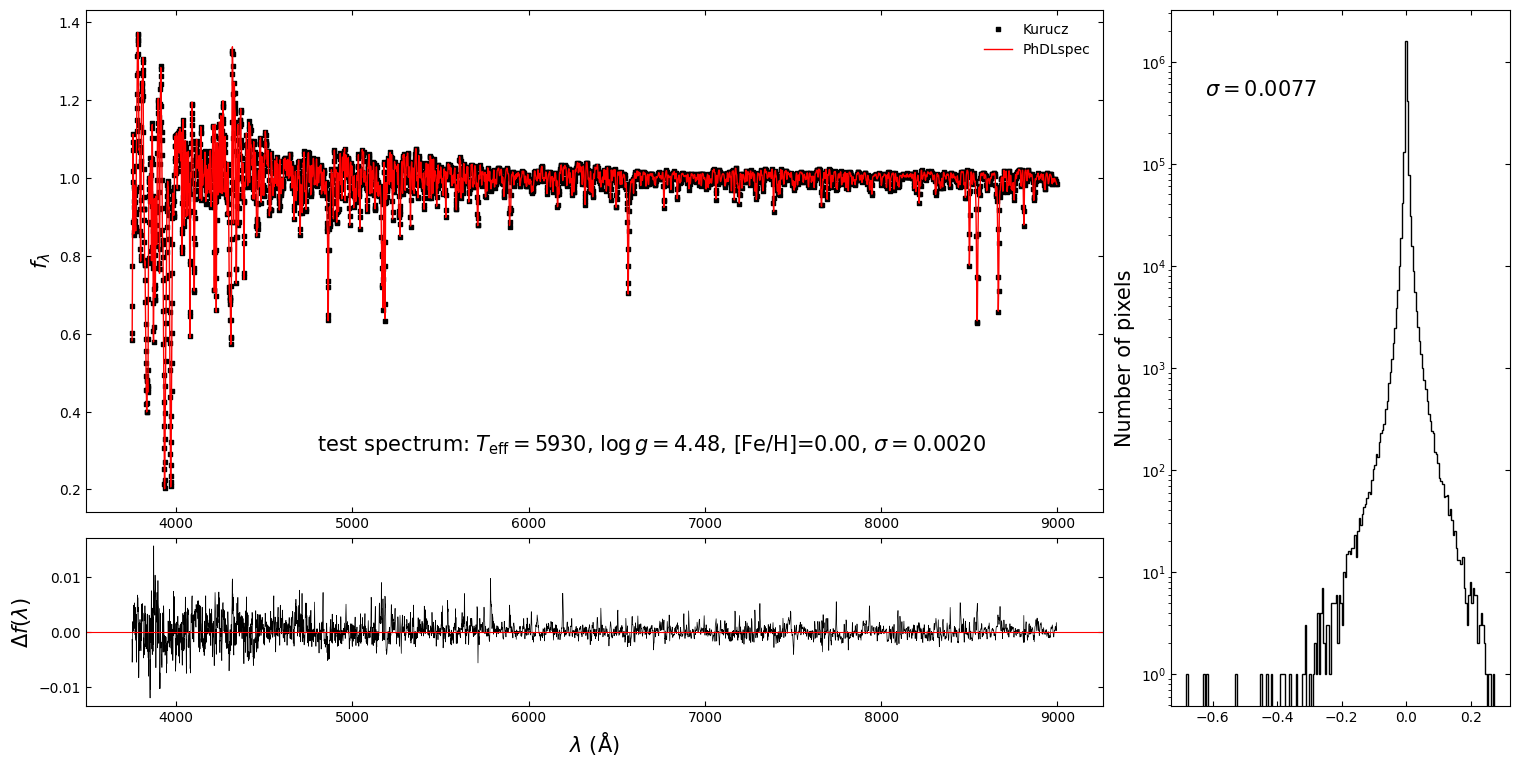}
		\caption{Comparison between the \texttt{Kurucz} spectrum (black squares) and the \texttt{PhDLspec} prediction (red line) for a test star with $T_{\rm eff}=5930$ K, $\log g=4.48$, and [Fe/H]=0.00 (left panel). The top subpanel shows the flux, and the bottom subpanel shows the residual ($\Delta f(\lambda) = f_{\rm PhDLspec} - f_{\rm Kurucz}$). The right panel displays histograms of the overall residuals for a test set of 623 spectra, each with 3800 pixels in wavelength. The standard deviation ($\sigma$) of the residuals is marked.}
       \label{fig:test_set_pixel_std}
\end{figure*}

The \texttt{Kurucz} spectra library was partitioned into training and test sets in a 9:1 ratio. We trained the \texttt{PhDLspec} model using the \texttt{Pytorch} framework on an NVIDIA GeForce RTX 4090 machine. To fit the memory, we divided the full spectrum into 11 segments, and trained the model segment by segment. Each segment was trained for 3000 epochs, utilizing a batch size of 256 and a cosine-annealing adjustable learning rate starting from 0.001. We adopted \texttt{Adam} \citep{2014arXiv1412.6980K} as the optimizer. Training was stopped if the test loss did not improve for 500 consecutive epochs. The model with the lowest test loss was retained. 

Figure \ref{fig:test_set_pixel_std} shows the residual of spectra emulation for the test set. The left panel shows an example comparison between a predicted spectrum (red line) and the corresponding \texttt{Kurucz} model (black squares) for a representative star with $T_{\rm eff}=5930$\,K, $\log g=4.48$, and $\text{[Fe/H]}=0.00$. The two spectra are almost indistinguishable, with a pixel-level dispersion of 0.2\%.
The right panel summarizes the residual distribution for the entire test set. 
For all pixels combined, the residual dispersion is around 0.77\%. A substantial fraction of this dispersion arises from cooler stars, whose spectra exhibit dense molecular features that are more challenging to model accurately. If we restrict the sample to spectra with effective temperatures above 4500 K, the dispersion decreases to 0.46\%.
These results demonstrate that \texttt{PhDLspec} achieves sub-percent precision in reproducing \texttt{Kurucz} spectra across a high-dimensional parameter space (over 30 dimensions), trained on a mere $\simeq$6000 model spectra.

\begin{figure*}[htbp]
		\centering
		\includegraphics[width=1\linewidth]{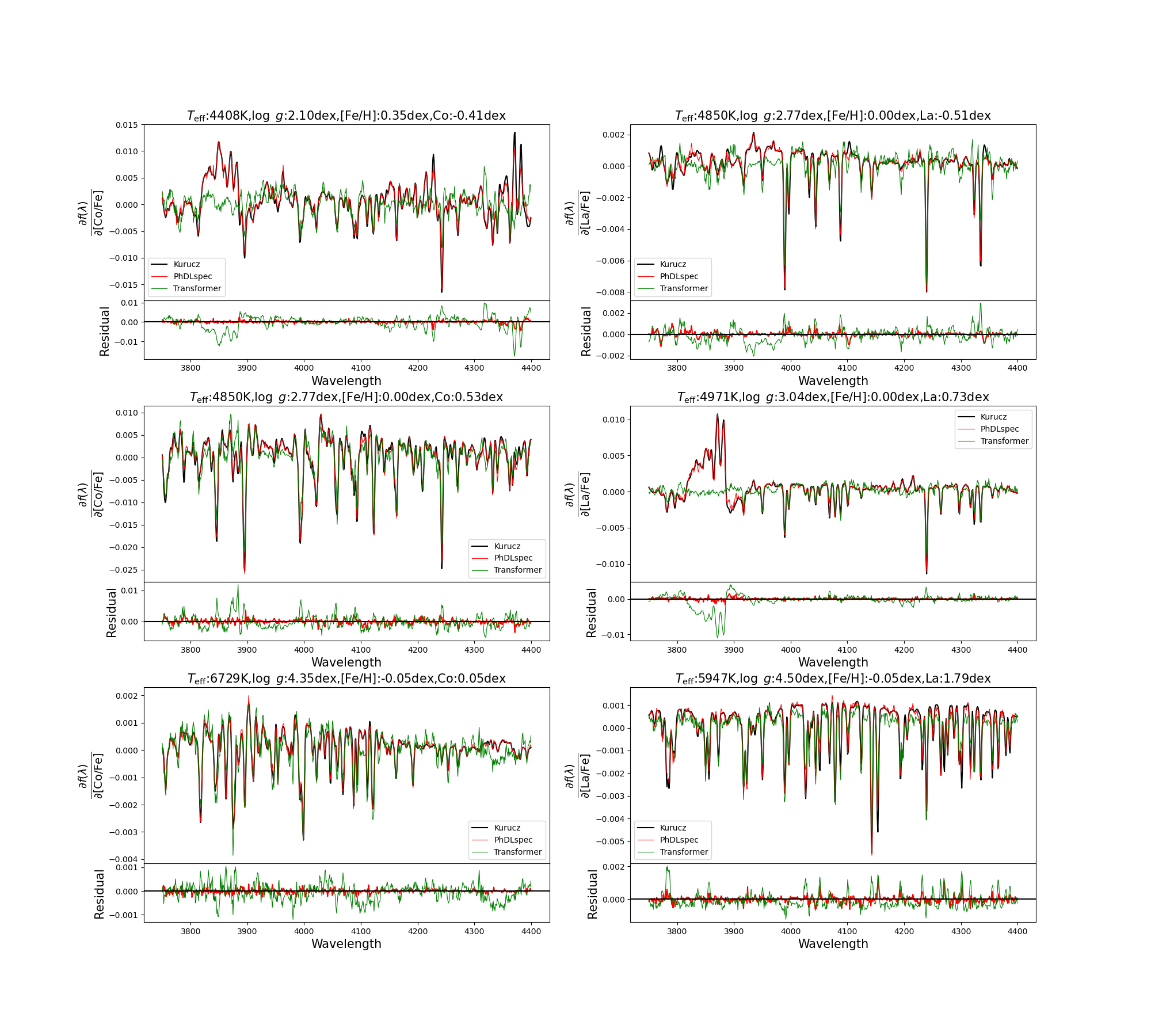}
		\caption{The partial derivatives of the normalized fluxes to elemental abundance ratios [Co/Fe] and [La/Fe] for a representative set of stars with different stellar atmospheric parameters ($T_{\rm eff}$, $\log g$, [Fe/H]). For each case, the upper subpanel shows the reference gradients from the \texttt{Kurucz} model (black) alongside the predictions from \texttt{PhDLspec} (red) and the Transformer model (green). The lower subpanel shows the residuals relative to the \texttt{Kurucz} spectra. Results for only wavelength window 3750--4400{\AA} are shown.}
       \label{fig:pred_spec}
\end{figure*}

To demonstrate the power of \texttt{PhDLspec} to recover physical spectral gradients, we also trained an ordinary Transformer-encoder model without imposing the physical priors term in the loss function. Figure~\ref{fig:pred_spec} shows a comparison of the emulated gradient spectra $\nabla_\theta{f(\lambda; \mathbf{\theta})}$ with that of the \textit{ab initio} \texttt{Kurucz} model computation, taking the cases of [Co/Fe] and [La/Fe] as examples. The comparison highlights clearly the advantage of \texttt{PhDLspec} over the ordinary Transformer model for recovering the ground-truth gradient spectra. As shown in the figure, \texttt{PhDLspec} accurately reproduces the \texttt{Kurucz} \textit{ab initio} gradient spectra across different stellar types, much better than the performance of the ordinary Transformer model. The advantage becomes especially pronounced for cooler stars and for elements with weak or broad spectral features.
This advantage ensures \texttt{PhDLspec}'s unique ability to determine stellar parameters in a rigorously physics-sensible way, which may not be available in ordinary Transformer model, even though the latter is trained on \textit{ab initio} model spectra. 


\subsection{Parameter inference for the test set}
\begin{figure*}[htbp]
		\centering
		\includegraphics[width=1\linewidth]{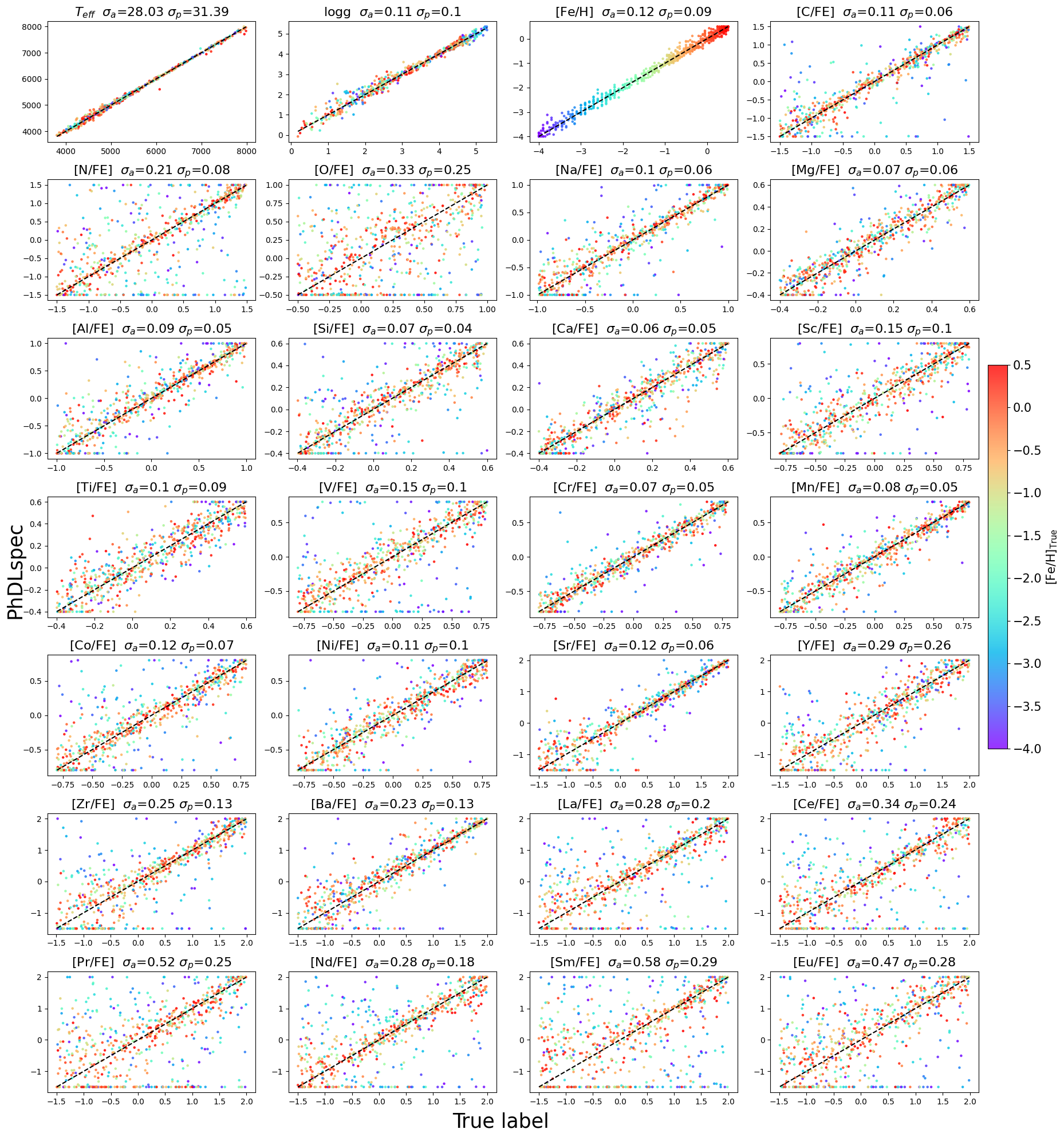}
		\caption{Comparison between the stellar labels derived with \texttt{PhDLspec} and the true labels for the test spectra set. Each panel shows one parameter ratio, with the one-to-one black dashed line indicating perfect agreement. The color represents the [Fe/H] value of the test spectra. The standard deviation of the residuals is presented both for the entire test set ($\sigma_{\mathrm{a}}$) and for metal-rich stars (${\rm [Fe/H]}>-1$) with $4500~\text{K}<T_{\rm eff}<7000~\text{K}$ ($\sigma_{\mathrm{p}}$). Overall, the \texttt{PhDLspec} measurements are consistent with the true labels, especially for metal-rich stars, with dispersion from $0.05$ to $0.3$~dex in the abundance differences, dependent on elements.}
       \label{fig:kurucz_compare}
\end{figure*}

After training the model, we employed the CMA-ES algorithm to fit the spectra of the test set, which consists of 623 \texttt{Kurucz} model spectra, after adding random noise corresponding to a median signal-to-noise ratio (S/N) of 100. 
Figure~\ref{fig:kurucz_compare} presents the one-to-one comparison between the best-fitting stellar parameters obtained by \texttt{PhDLspec} and the true values, color-coded by [Fe/H]. 
Overall, the fitted values exhibit an excellent correlation with the true labels, with a standard deviation of 28.03~K in $T_{\rm eff}$, 0.11~dex in $\log g$, 0.12~dex in [Fe/H]. The standard deviation in abundance ratios [X/Fe] exhibits a significant variation across $T_{\rm eff}$, [Fe/H], and [X/Fe] themselves. For metal-rich stars (${\rm [Fe/H]}>-1$) with $4500<T_{\rm eff}<7000$~K, the standard deviation is smaller than 0.1~dex for [C/Fe], [N/Fe], [Na/Fe], [Mg/Fe], [Al/Fe], [Si/Fe], [Ca/Fe], [Sc/Fe], [Ti/Fe], [V/Fe], [Cr/Fe], [Mn/Fe], [Co/Fe], [Ni/Fe], [Sr/Fe], 0.1-0.2~dex for [Zr/Fe], [Ba/Fe], [La/Fe], [Nd/Fe], 0.2-0.3~dex for the remaining elements. For metal-poor stars or stars with low [X/Fe], the standard deviation becomes dramatically larger, owing to the dramatic diminishment of spectra gradients in the spectra. 



\subsection{The correlation matrix}

The gradient spectra $\nabla_\theta{f(\lambda; \mathbf{\theta})}$ for different stellar parameters are non-orthogonal. In low-resolution spectra, line blending further reduces this orthogonality, which increases the off-diagonal terms of the covariance matrix (Equation~\ref{eq:covmatrix}) and leads to significant correlations between parameter estimates.

\begin{figure*}[htbp]
		\centering
		\includegraphics[width=1\linewidth]{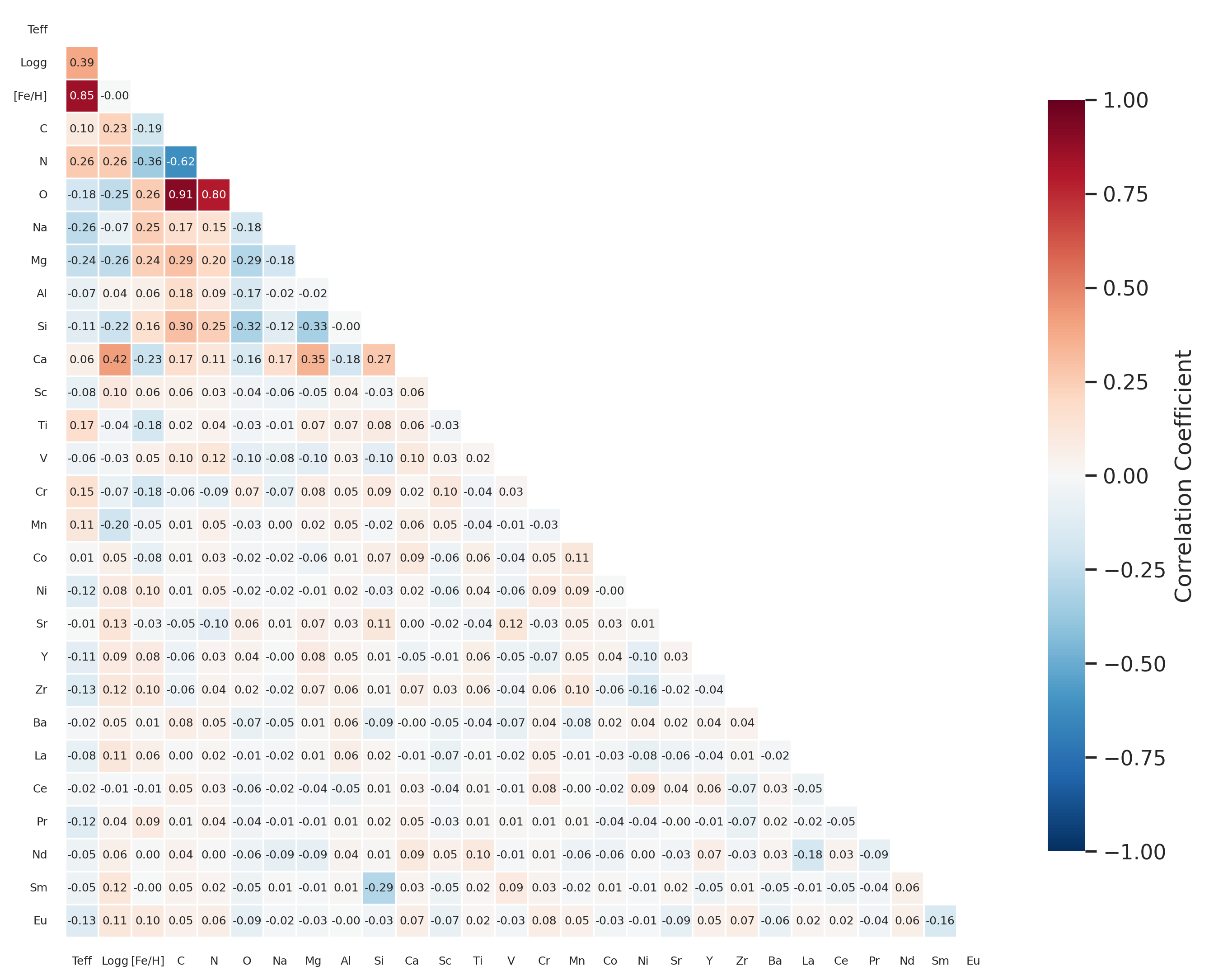}
		\caption{Correlation matrix between stellar atmospheric parameters ($T_{\mathrm{eff}}$, $\log g$, and [Fe/H]) and abundances of 25 elements, derived by computing the Pearson correlation coefficient of their gradient spectra. Strong correlations are seen between [Fe/H] and many elemental abundances, as well as among the light elements (e.g., O and N).}
       \label{fig:covmatrix}
\end{figure*}

Figure~\ref{fig:covmatrix} shows the correlation matrix derived from the gradient spectra of a fiducial star with $T_{\rm eff}=4800$\,K, $\log g=2.5$, and [Fe/H]=0.0. The matrix reveals strong correlations among parameters whose features are prominent across many wavelengths, such as $T_{\rm eff}$, $\log g$, [Fe/H], [C/Fe], [N/Fe], and [O/Fe]. In contrast, correlations between abundances of heavy elements are negligible, consistent with previous studies \citep{2017ApJ...849L...9T, 2019ApJS..245...34X, 2025ApJS..279....5Z}. The strong C-N-O correlation arises from the CNO atomic-molecular network, as demonstrated by \citet{2018ApJ...860..159T}. We also find moderate-to-strong correlations between [Ca/Fe], [Mg/Fe], and other parameters like [Si/Fe], [Fe/H], and $\log g$, primarily due to the sensitivity of strong features like the Ca~{\sc ii} H and K lines and the Mg~{\sc i} b triplet to the spectral gradients of these parameters.

\section{External calibration and validation}\label{sec:calibration}
It turns out that stellar parameters determined based on \textit{ab initio} model spectra suffer strong systematic errors as a consequence of imperfections in the model spectra. We therefore implement dedicated external calibration and validation for the estimated parameters, using a number of calibration sources, including 
\begin{itemize}
\item common stars among LAMOST, APOGEE, and GALAH, which are used for calibration of basic parameters ($T_{\rm eff}$, $\log g$, [Fe/H]; section 4.1), chemical abundances (section 4.2.2), and validation (section 4.4);
\item wide binaries identified with Gaia astrometry, which are used for internal calibration of the abundance determinations (section 4.2.1, section 4.2.2);  
\item member stars of open cluster M67, which are used for abundance calibration (section 4.2.2) and validation of chemical abundances (section 4.3);
\item abundances from high-resolution spectroscopy in literature (section 4.2.2).
\end{itemize}
 
\subsection{Calibration of stellar atmospheric parameters}
The stellar atmospheric parameters are compared and calibrated to the \texttt{DD-Payne} estimates from LAMOST spectra \citep{2025ApJS..279....5Z}. Effective temperature of the LAMOST \texttt{DD-Payne} estimates are calibrated to the infrared flux method (IRFM) scale of \citet{2009A&A...497..497G}. Surface gravity of the LAMOST \texttt{DD-Payne} estimates inherit the scale of the APOGEE survey, which is adopted as the training set, and show good consistency with asteroseismic measurements (see Figure 13 of \citet{2025ApJS..279....5Z}). The LAMOST \texttt{DD-Payne} catalog provides two sets of [Fe/H] estimates, taking high-resolution spectroscopy from the GALAH and APOGEE, respectively, as the training sets. Here we adopt the former, as it is corrected for non-local equilibrium effect (NLTE). 

To implement the calibration, we first build a calibration sample by cross-matching LAMOST DR10 with both GALAH DR3 \citep{2021MNRAS.506..150B} and APOGEE DR17 \citep{2022ApJS..259...35A}, yielding 3,283 common stars among these three surveys. We further apply a signal-to-noise ratio ($S/N$) criteria,
\[
\begin{cases}
S/N\_{\rm c1}\_{\rm iraf} > 40, \\
S/N\_{\rm c3}\_{\rm iraf} > 60, \\
S/N_g > 40,
\end{cases}
\]
where $S/N\_{c1}\_{iraf}$ and $S/N\_{c3}\_{iraf}$ are spectral $S/N$ of GALAH spectra, $S/N_g$ is spectral $S/N$ of LAMOST spectra in SDSS $g$ band. These criteria result in a high-quality sample of 440 stars. These common sources will be used later to assess the consistency between the \texttt{PhDLspec} method and high-resolution spectroscopic survey results (see Section~\ref{sec:high-resolution results} for details), but at this stage, they are employed for calibration of atmospheric parameters, taking their LAMOST \texttt{DD-Payne} estimates in \citet{2025ApJS..279....5Z} as the reference. 

\begin{figure}[!htbp]
		\centering
		\includegraphics[width=1\linewidth]{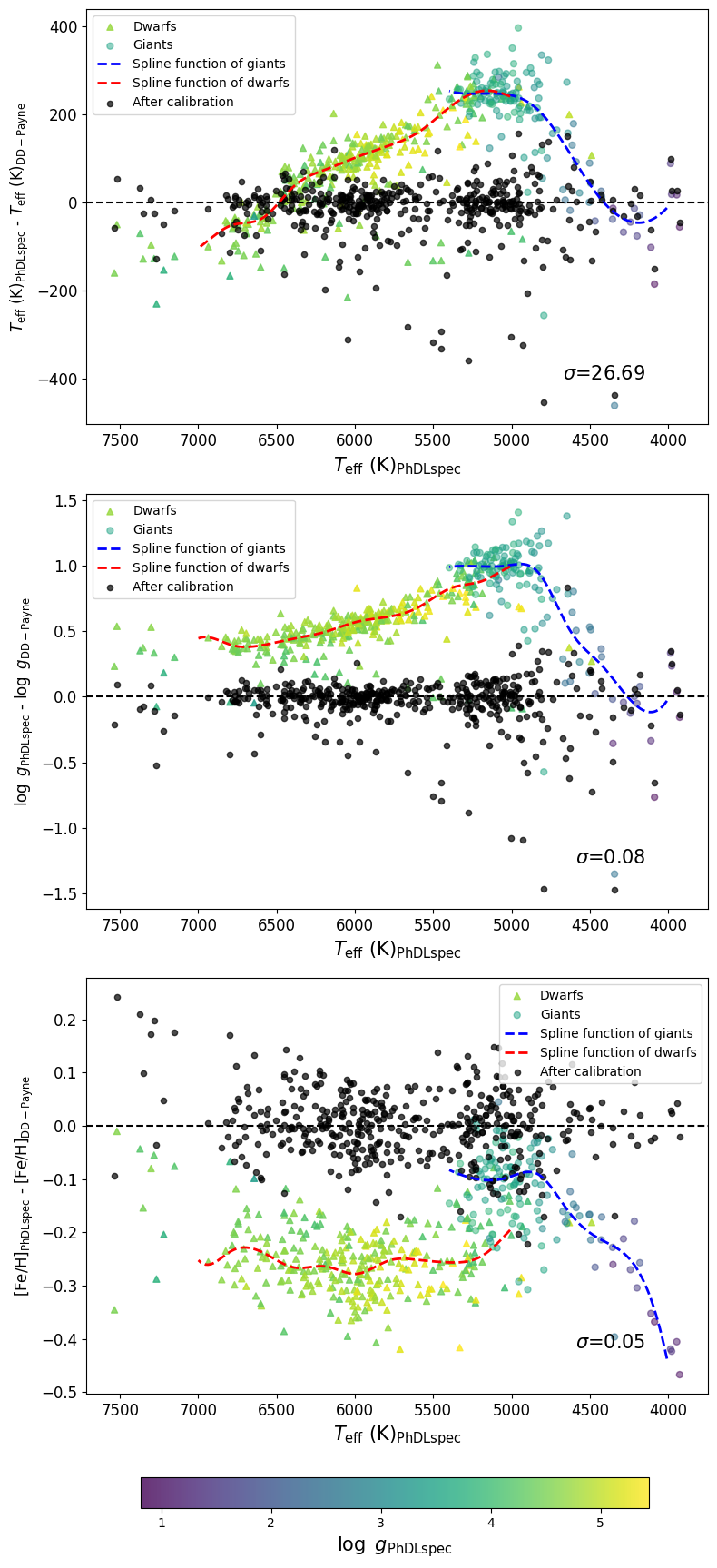}
		\caption{Comparison of the atmospheric parameters derived with \texttt{PhDLspec} and those from the \texttt{DD-Payne} estimates of \citet{2025ApJS..279....5Z}. From top to bottom, the residuals of $T_{\rm eff}$, $\log g$, and [Fe/H] as a function of $T_{\rm eff}$ from \texttt{PhDLspec} are presented, respectively. Triangles and circles denote dwarf and giant stars, respectively, with colors representing their $\log g$ values. Red and blue dashed lines show cubic spline fits for dwarfs and giants, respectively. Black dots show results after removing the systematic trends from the cubic spline fits. The standard deviation ($\sigma$) in each panel quantifies the scatter of the residuals after calibration.}
       \label{fig:atmos}
\end{figure}

Figure~\ref{fig:atmos} shows the differences in the atmospheric parameters between our \texttt{PhDLspec} results and the \texttt{DD-Payne} estimates. We apply cubic spline interpolation to determine the median differences as a function of $T_{\rm eff}$, and subtract them from the \texttt{PhDLspec} results, for dwarfs and giants ($T_{\rm eff}<5400$~K, $\log g<4$~dex), respectively. For stars outside the interpolation range, we apply fixed boundary corrections: giants with $T_{\rm eff}<4000$ K use the correction at 4000 K, and dwarfs with $T_{\rm eff}<5000$ K or $T_{\rm eff}>7000$ K use the corrections at 5000 K and 7000 K, respectively.
The residuals after correcting for the mean trends exhibit a dispersion of 26.7~K in $T_{\rm eff}$, 0.08~dex in $\log g$, and 0.05~dex in [Fe/H]. 

\subsection{Calibration of elemental abundances}

\begin{figure*}[htbp]
		\centering
		\includegraphics[width=0.9\linewidth]{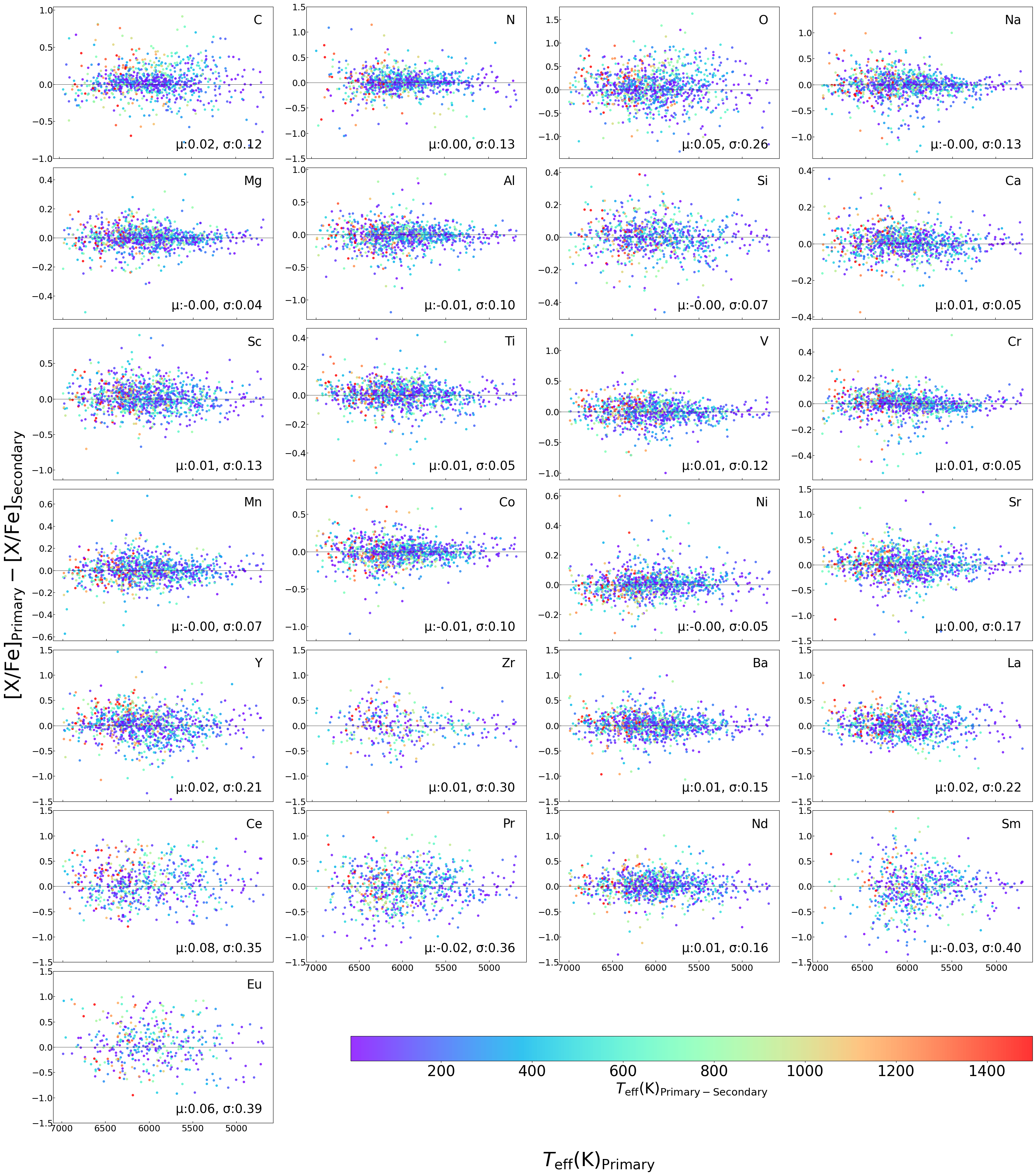}
		\caption{Abundance differences between the primary and secondary stars of wide binary systems as a function of the primary’s effective temperature. Each panel shows \(\text{[X/Fe]}_{\rm Primary} - \text{[X/Fe]}_{\rm Secondary}\) for a given element. 
		The color bar represents the effective temperature difference between the primary and secondary. Median and standard deviation of the differences are marked in each panel.}
       \label{fig:widebinary}
\end{figure*}

Systematic trends in the derived elemental abundances are observed with respect to both $T_{\rm eff}$ and $\log g$. To mitigate these effects and establish a consistent scale with external measurements, we performed an empirical calibration. In doing so, we assume the systematic error $\epsilon$ can be decomposed into a term that depends only on temperature and another that depends only on $\log g$, as expressed below
\begin{equation}
\begin{aligned}
\overline{\mathrm{[X/Fe]}}_\texttt{PhDLspec}&=\overline{\mathrm{[X/Fe]}}_{\rm true}+\epsilon\\
\epsilon&=f(T_{\rm{eff}})+f(\log g)+d.
\label{eq:t}
\end{aligned}
\end{equation}
These systematic error components are then determined in two steps using independent samples. In the first step, we determine and correct for temperature-dependent trends using wide binaries, assuming that component stars of a binary system share the same intrinsic chemical composition.  
In the second step, we correct for systematic trends with $\log g$ by comparing the \texttt{PhDLspec} abundances with high-resolution spectroscopic measurements from the Hypatia catalog \citep{2014AJ....148...54H}. The constant term $d$ in Equation \ref{eq:t} represents the global zero-point offset, which is subsequently constrained through an independent calibration based on the absolute zero-points from high-resolution surveys.

\subsubsection{Correcting for $T_{\rm{eff}}$-dependent trend of [X/Fe] using wide binaries}
We cross-matched the LAMOST catalog with the Gaia wide binary catalog of \citet{2021MNRAS.506.2269E}, and identified wide binaries that both components have LAMOST observations and satisfy the following criteria,
\[
\left\{
\begin{aligned}
& S/N > 40, \\
& 4500~\text{K} < T_{\rm eff} < 7000~\text{K}, \\
& \log g > 4, \\
& \Delta{V_{\rm los}}/{\mathrm {mad}}(\Delta{V_{\rm los}}) < 3, 
\end{aligned}
\right.
\]
where ${\mathrm {mad}}(\Delta{V_{\rm los}})$ denotes the median absolute deviation of the line-of-sight velocity difference between the binary components. These criteria yield a sample of 1,168 wide binary systems used for the calibration. 
\begin{table}[ht]
\centering
\caption{The polynomial coefficients ($a$, $b$, $c$) for the temperature-dependent systematic error term are derived from wide binaries, with an additional constant term $d$ obtained from zero-point correction.}
\begin{tabular}{ccccc}
\toprule
Element([X/Fe]) & a & b & c & d \\
\midrule
C  & -0.03 & 0.40 & -1.94 & 3.04 \\
N  & 0.34  & -5.54 & 29.75 & -51.21 \\
O  & -0.03 & 0.64 & -4.63 & 0.05 \\
Na & 0.04  & -0.48 & 1.61  & 0.08 \\
Mg & 0.01  & -0.28 & 1.72  & -0.005 \\
Al & -0.05 & 0.77  & -4.33 & 0.05 \\
Si & -0.01 & 0.15  & -0.97 & 0.003 \\
Ca & -0.04 & 0.78  & -4.96 & -0.034 \\
Sc & 0.00  & 0.08  & -1.30 & -0.06 \\
Ti & -0.00 & 0.06  & -0.42 & 0.0028 \\
V  & 0.05  & -0.97 & 5.92  & 0.0087 \\
Cr & -0.01 & 0.16  & -0.92 & 0.005 \\
Mn & -0.04 & 0.71  & -4.40 & 0.024 \\
Co & -0.10 & 1.74  & -10.18 & -0.007 \\
Ni & -0.09 & 1.51  & -8.45 & 0.026 \\
Sr & 0.06  & -0.82 & 3.07  & -0.014 \\
Y  & 0.20  & -3.30 & 17.93 & -0.04 \\
Zr & 0.21  & -3.58 & 20.19 & -0.12 \\
Ba & 0.11  & -1.80 & 9.30  & 0.01 \\
La & 0.09  & -1.88 & 12.68 & -0.3 \\
Ce & -0.07 & 1.20  & -6.42 & -0.06 \\
Pr & -0.02 & 0.15  & -0.20 & -0.018 \\
Nd & 0.16  & -2.89 & 16.46 & -0.32 \\
Sm & 0.27  & -4.65 & 26.18 & -0.14 \\
Eu & -0.25 & 3.96  & -21.07 & -0.22 \\
\bottomrule
\end{tabular}
\label{tab:elements_coefficients}
\end{table}

We assume that the dependence of chemical abundance on $T_{\rm{eff}}$ can be described by a third-order polynomial,
\begin{equation}
\begin{aligned}
f(T_{\rm{eff}})=aT_{\rm{eff}}^{3}+bT_{\rm{eff}}^{2}+cT_{\rm{eff}}.
\label{eq:fteff}
\end{aligned}
\end{equation}
where the three coefficients $a$, $b$, and $c$ are constrained using wide binaries. The constant term is not specified here and will be fixed during the zero-point calibration process.

The abundance difference between the binary components can thus be expressed as:
\begin{equation}
\begin{aligned}
\Delta\text{[X/Fe]}&=a(T_{\rm{eff},1}^{3}-T_{\rm{eff},2}^{3})\\&+b(T_{\rm{eff},1}^{2}-T_{\rm{eff},2}^{2})\\&+c(T_{\rm{eff},1}-T_{\rm{eff},2}).
\label{eq:calibration}
\end{aligned}
\end{equation}
The coefficients $a$, $b$, and $c$ were determined by fitting this relation to the binary sample using a Markov Chain Monte Carlo (MCMC) approach implemented with the \texttt{emcee} package \citep{2013PASP..125..306F} in Python. The resulting coefficients for different elements are listed in Table~\ref{tab:elements_coefficients}. 

Figure~\ref{fig:widebinary} presents the abundance differences between the components of wide binaries after applying the effective temperature calibration, plotted as a function of the primary star’s effective temperature and color-coded by the temperature difference between the two stars. Compared to the uncalibrated results (Appendix, Figure~\ref{fig:widebinary_trend}), the calibration has significantly reduced the spurious trends with $T_{\rm eff}$ for several elements, including N, Mg, Ni, Sc, Sr, Nd, La, and Pr. This correction provides a more faithful representation of the intrinsic stellar abundances and is expected to improve the reliability of subsequent analyses, such as chemical tagging and cross-comparison of stars across a broad temperature range.

\subsubsection{Calibrating abundance trend with $\log g$}
While the calibration based on wide binaries effectively removes systematic trends with effective temperature, it turns out that the abundances still exhibit an artificial trend with surface gravity. To achieve better calibration, we constructed a calibration sample incorporating previously used wide binaries alongside common sources from LAMOST, APOGEE, and GALAH. The M67 open cluster was also included in the calibration sample. After retaining only the highest signal-to-noise ratio unique matching entries, we obtained a refined subset of 179 member stars. These stars are also used to validate the consistency of the derived abundances in Section~\ref{sec:starcluster}.
Due to the lack of giant stars in the sample, we further supplement another 200 giant stars with solar metallicity from the LAMOST–APOGEE common sources that meet the following criteria:
\[
\left\{
\begin{aligned}
&\text{$S/N_g$} > 50, \\
&\text{$\log g$} <3.2, \\
&-0.1 < \mathrm{[Fe/H]} < 0.1.
\end{aligned}
\right.
\]
The [X/Fe] trends with (uncalibrated) $\log g$ of this combined stellar sample are then compared to the high-resolution spectroscopic abundances from the Hypatia catalog \citep{2014AJ....148...54H}, a homogenized compilation of elemental abundances for nearby FGKM-type stars. 

\begin{table}[htbp]
\centering
  \caption{The absolute zero-points from the high-resolution survey used for external calibration.}
  \label{tab:zeropoints}
    \begin{tabular}{ccc}
\toprule
Element([X/Fe]) & Zero points   &Source  \\ \midrule
C &$-$0.01&Hypatia\\
N &$-$0.05 &Hypatia \\
O&  $-$0.05&Hypatia\\
Na&  $-$0.11&Hypatia\\
Mg&   0.00&Hypatia\\
Al&  $-$0.06&Hypatia\\
Si&  $-$0.01&Hypatia \\
Ca&  0.03&Hypatia\\
Sc&  0.06&Hypatia\\
Ti&  0.00 &Hypatia\\
V&  $-$0.02&Hypatia\\
Cr & $-$0.01&Hypatia\\
Mn& $-$0.03&GALAH\\
Co&  0.00&Hypatia\\
Ni&  $-$0.02&Hypatia\\
Sr& 0.01&Hypatia\\
Y& 0.03&Hypatia\\
Zr&  0.11&Hypatia\\
Ba&  0.00 &GALAH\\
La& 0.26&GALAH\\
Ce&  0.04&GALAH\\
Pr&   0.00&/\\
Nd & 0.30&GALAH\\
Sm & 0.10&GALAH\\
Eu&  0.19&GALAH\\
\bottomrule
    \end{tabular}
\end{table}

\begin{figure*}[htbp]
		\centering
        \includegraphics[width=0.9\linewidth]{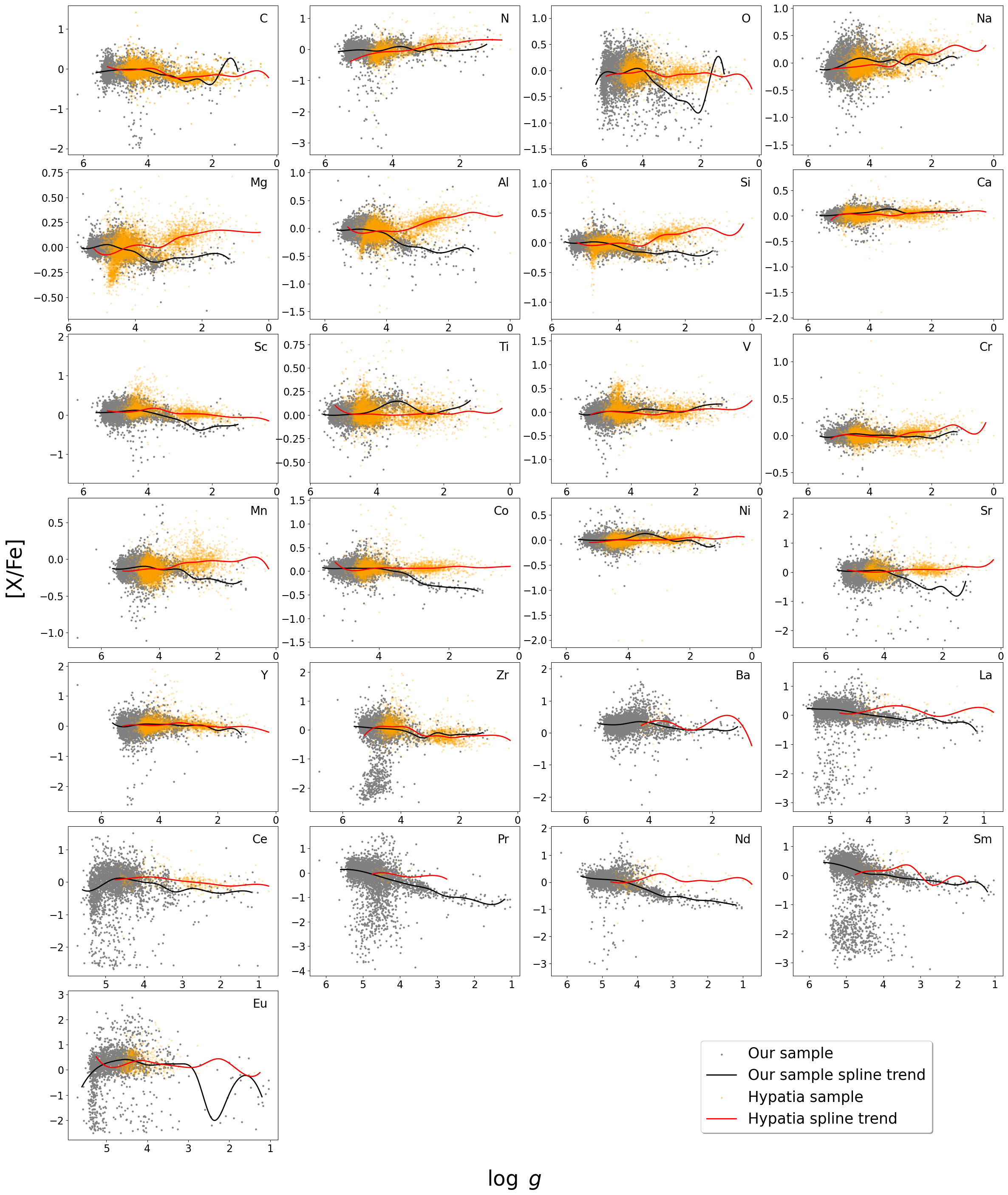}
		\caption{Elemental abundance ratios [X/Fe] as a function of surface gravity. Gray dots represent \texttt{PhDLspec} abundance estimates from LAMOST spectra for our sample stars with ${\rm [Fe/H]}>-0.5$, while orange dots are high-resolution spectroscopic measurements compiled in the Hypatia Catalog \citep{2014AJ....148...54H} for stars with ${\rm [Fe/H]}>-0.2$. Black and red lines indicate cubic spline fits to the binned medians for our sample and the Hypatia sample, respectively.}
       \label{fig:logg_trend_correct}
\end{figure*}

Such a comparison is illustrated in Figure~\ref{fig:logg_trend_correct}, where we confined the comparison to stars with solar metallicity by adopting a cut of  ${\rm [Fe/H]}>-0.5$ in our estimates, and ${\rm [Fe/H]}>-0.2$ in the Hypatia catalog. Again, we adopt a cubic spline function to determine the mean trends of abundances as a function of $\log g$. The Figure shows that for most elements, abundances in the Hypatia catalog exhibit a flat trend that is independent of $\log g$, whereas for many of them such [O/Fe], [Mg/Fe], [Al/Fe], [Si/Fe], [Mn/Fe], [Co/Fe], [Sr/Fe], and [Nd/Fe], our measurements exhibit clear $\log g$-dependence trends, likely due to systematic errors in our results. These $\log g$-dependence trends are subtracted from our measurements to achieve robust abundance estimates. Note that the stellar surface abundance of C and N can be altered as a consequence of stellar evolution, which may also induce a $\log g$-dependence trend. We therefore opt not to implement any correction to [C/Fe] and [N/Fe] to avoid overcorrection. 

Here we also calibrated the abundance zero points, corresponding to the coefficient $d$ in Equation~\ref{eq:t}. To determine the reference values, we selected solar-twin stars from the Hypatia catalog satisfying $5700~\text{K}<T_{\rm  eff}<5900~\text{K}$, $4<\log g<5$, and $-0.1<\text{[Fe/H]}<0.1$, and computed the median abundance of each element. For elements with insufficient coverage in this catalog, we adopted zero points from GALAH DR3 or set them to zero when no reliable external reference was available. The adopted zero point values are summarized in Table~\ref{tab:zeropoints}.

Once the systematic error term $\epsilon$ is derived, the calibrated abundances can be obtained by subtracting $\epsilon$ from the measured values, $\text{[X/Fe]}_\texttt{PhDLspec}$.

\subsection{Validation with M67 member stars}\label{sec:starcluster}
After applying the two-step calibration procedure, this section presents the elemental abundance results for M67 member stars, as a validation of the calibration. 

\begin{figure*}[htbp]
		\centering
        \includegraphics[width=0.9\linewidth]{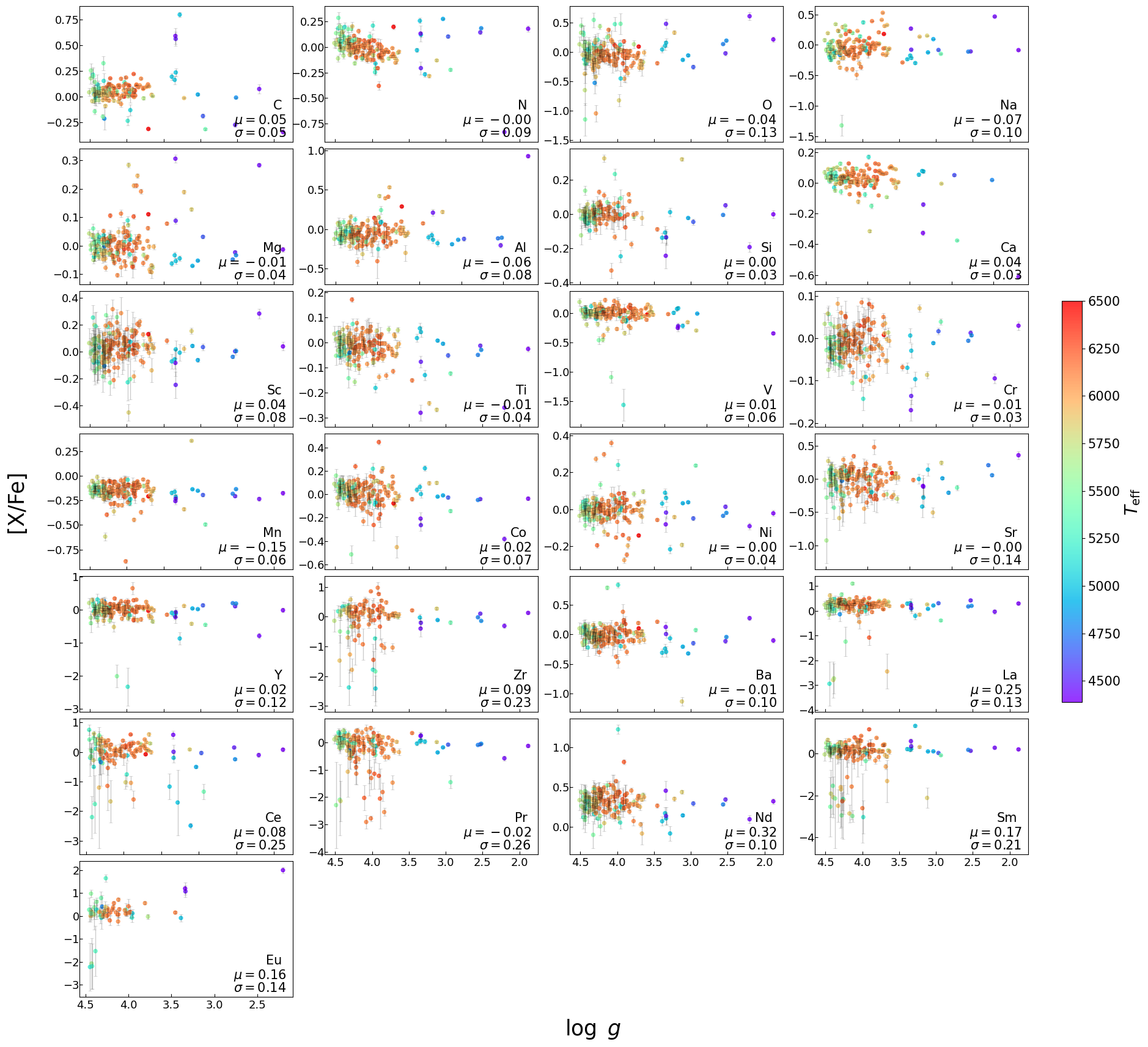}
		\caption{Abundance ratios [X/Fe] as a function of surface gravity $\log g$ for M67 member stars. Each panel corresponds to a different element, as indicated. The dots are color-coded by effective temperature $T_{\rm eff}$. The mean abundance ($\mu$) and dispersion ($\sigma$) are marked in the bottom-right corner of each panel.}
       \label{fig:m67_cali}
\end{figure*}

Figure~\ref{fig:m67_cali} shows the abundance ratios [X/Fe] of various elements as a function of $\log g$. For most elements except for N, the abundance ratios display nearly flat trends across different evolutionary stages from the main sequence to the red giant branch, with little dependence on $\log g$. The star-to-star scatter in [X/Fe], defined as the standard deviation after eliminating outliers, is only 0.03--0.04~dex for Mg, Si, Ca, Ti, Cr, and Ni, 0.05--0.10~dex for C, N, Na, Al, Sc, V, Mn, Co, Ba, and Nd, and 0.10--0.15~dex for O, Sr, Y, La, and Eu, and 0.20-0.30~dex for Zr, Ce, Pr, and Sm. These small values suggest a high degree of internal measurement precision of the \texttt{PhDLspec} on LAMOST spectra.   

\begin{figure*}[htbp]
		\centering
		\includegraphics[width=0.8\linewidth]{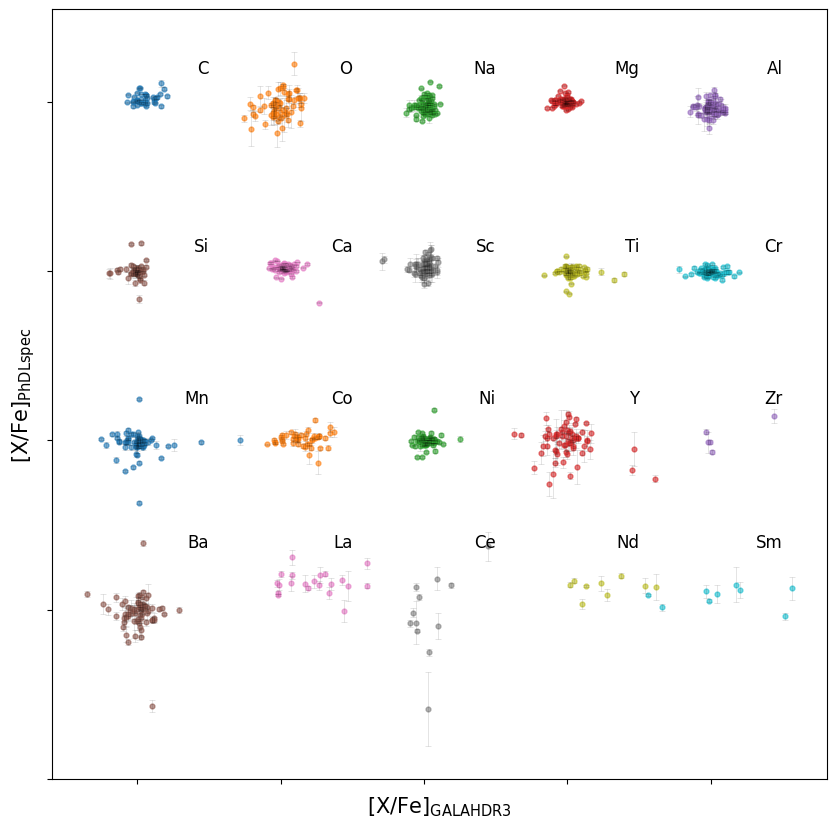}
		\caption{Comparison between the elemental abundances of M67 member stars derived with \texttt{PhDLspec} from LAMOST spectra and those from the GALAH DR3. The X- and Y-axes have the same scale unit, but are arbitrarily shifted for different elements. The scatter points represent individual stars, and error bars (when visible) indicate the measurement uncertainties.} 
       \label{fig:compare_m67_galah}
\end{figure*}

Nevertheless, Figure~\ref{fig:m67_cali} shows some substantial portion of outliers whose abundance ratios are significantly lower than the average value for some heavy elements, such as Zr, Ce, Pr, Sm, and Eu. The nature of these outlier measurements is unclear, but we mention that they have abnormally low values and much larger measurement errors than typical values. 

For an external comparison with elemental abundances derived from high-resolution spectra, we cross-matched our M67 member star sample with GALAH DR3, obtaining 77 stars in common. Figure \ref{fig:compare_m67_galah} displays the star-to-star scatter in both measurements for 20 elements in a single panel.  
Notably, the Figure exhibits an elongated distribution for many elements such as C, Ca, Ti, Cr, Mn, Co, Ni, La, Nd, and Sm, as the \texttt{PhDLspec} measurements from LAMOST spectra exhibit smaller abundance dispersion than the GALAH DR3 results. This confirms the ability of \texttt{PhDLspec} for delivering precise abundances from LAMOST low-resolution spectra. 


\subsection{Validation with high-resolution spectroscopic surveys} \label{sec:high-resolution results}
After applying the above calibrations, we evaluate the performance of \texttt{PhDLspec} by comparing the calibrated parameters with those of high-resolution survey results from both APOGEE DR17 \citep{2022ApJS..259...35A} and GALAH DR3 \citep{2021MNRAS.506..150B} for common stars.

\begin{figure*}[htbp]
		\centering
		\includegraphics[width=1\linewidth]{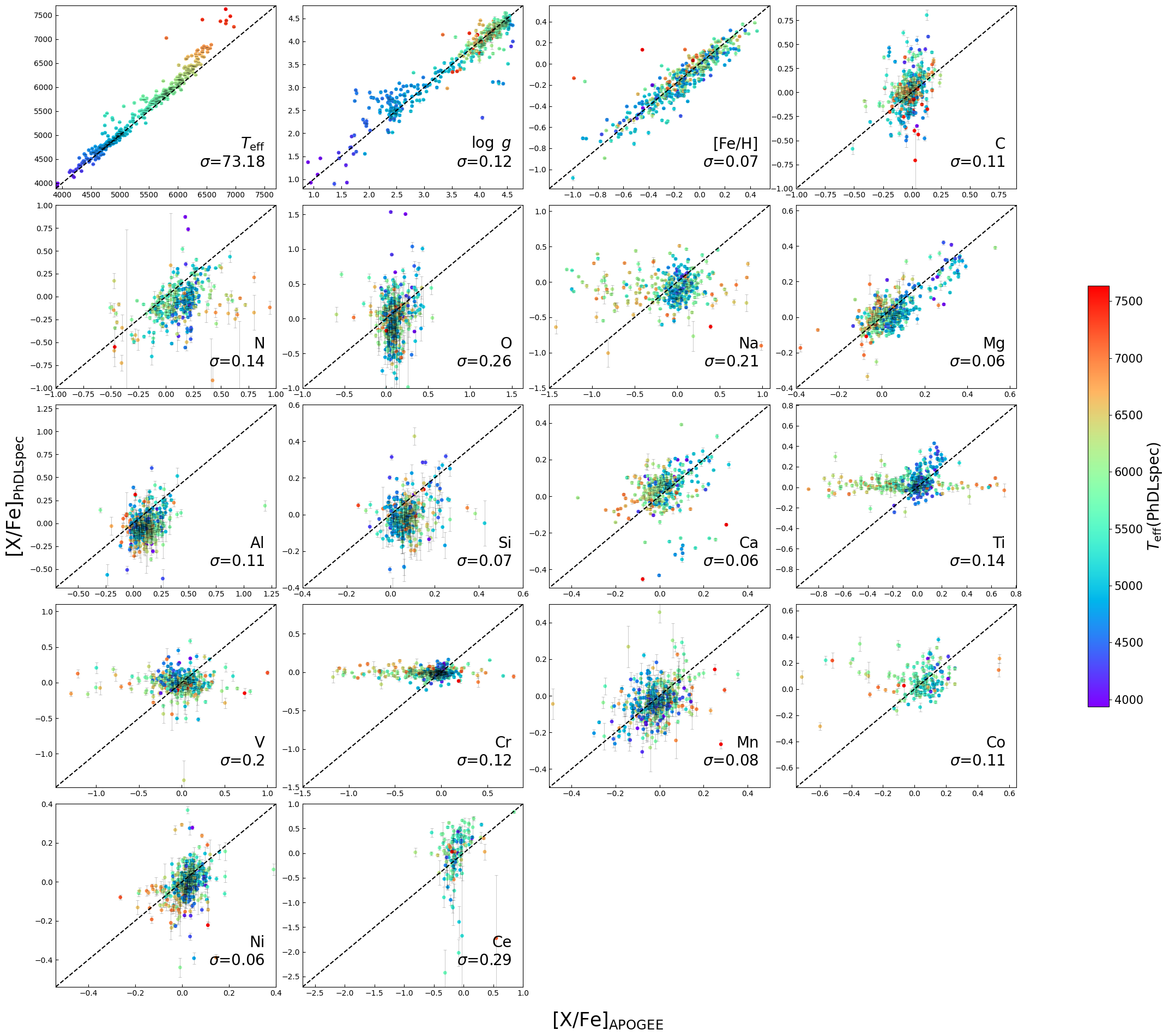}
		\caption{One-to-one comparison of the atmospheric parameters and elemental abundances ([X/Fe]) derived with \texttt{PhDLspec} from LAMOST spectra with those from APOGEE DR17 \citep{2022ApJS..259...35A} for stars in common. Each panel shows the comparison for a label, with the dashed line indicating the 1:1 relation. Points are color-coded by $T_{\rm eff}$ derived with \texttt{PhDLspec}. The standard deviation ($\sigma$) of the label differences is marked in each panel.}
       \label{fig:lamapo2}
\end{figure*}

Figure~\ref{fig:lamapo2} shows the comparison between the LAMOST \texttt{PhDLspec} parameters after calibration and APOGEE. The difference in $T_{\rm eff}$ exhibits a dispersion of 73~K. This value is partly contributed by a systematic deviation at the regime of $T_{\rm eff}>6000$~K, where the APOGEE DR17 temperature are underestimated \citep{2025ApJS..279....5Z}. The $\log g$ and [Fe/H] show good agreement in general, and the overall differences have a dispersion of 0.12~dex and 0.07~dex, respectively. However, a small port of giant stars at $\log g\simeq3$ show pronounced deviation in $\log g$ from their APOGEE values, likely a consequence of imperfect calibration of the \texttt{PhDLspec} estimates. The [X/Fe] shows good agreement within the measurement errors for a number of elements, including C, O, Mg, Al, Si, Ca, Mn, and Ni. For other elements especially N, Na, Ti, B, Cr, and Co, the Figure exhibits a significant portion of stars whose abundance ratios are normal in the LAMOST \texttt{PhDLspec} results, but are either too high or too low in the APOGEE estimates, probably attribute to errors in the APOGEE measurements (see also similar findings in \cite{2025ApJS..279....5Z}). 


\begin{figure*}[htbp]
		\centering
		\includegraphics[width=1\linewidth]{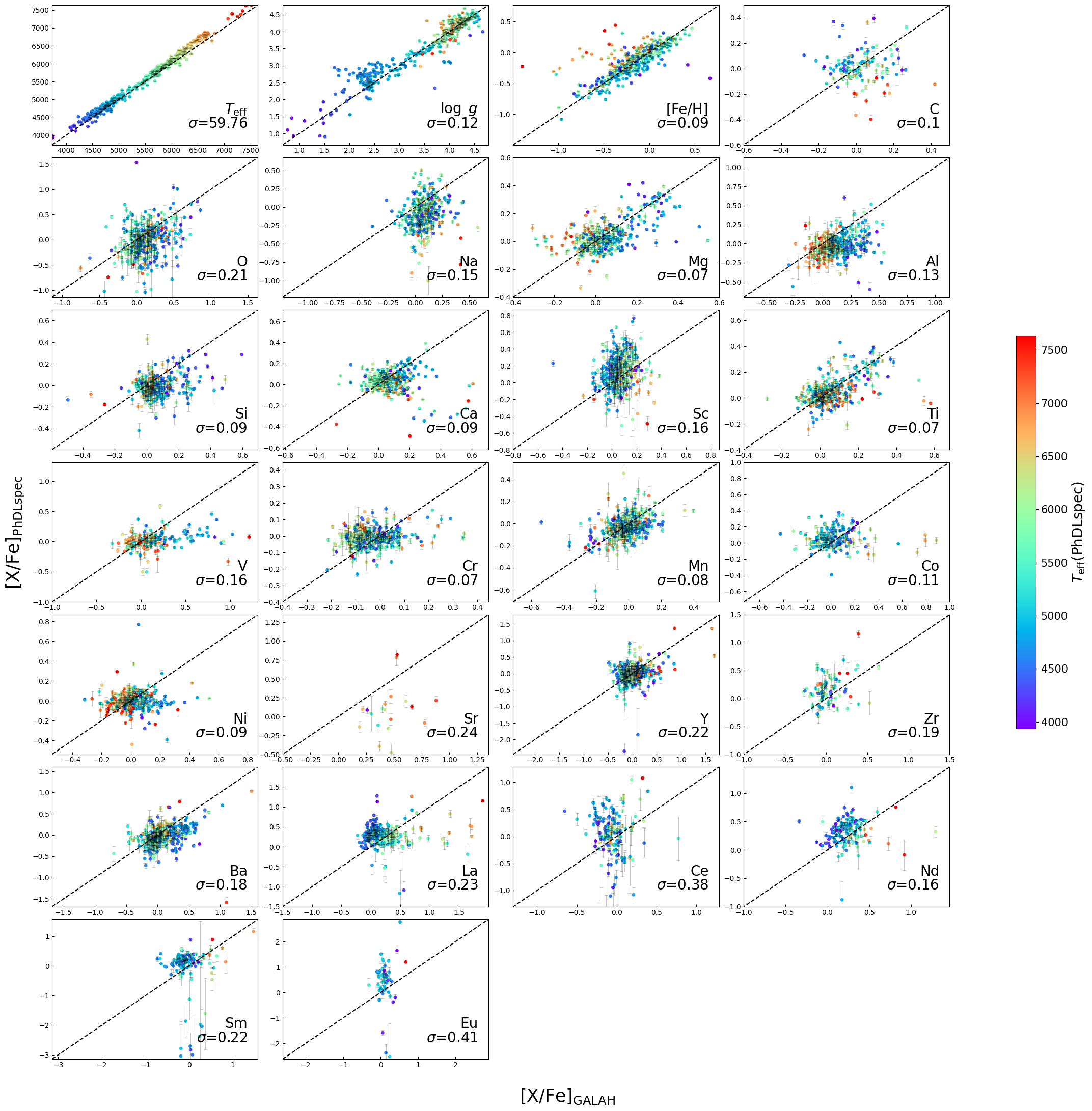}
		\caption{Same as Figure~\ref{fig:lamapo2}, but for comparison with GALAH DR3.}
       \label{fig:lamgalah2}
\end{figure*}

Figure~\ref{fig:lamgalah2} shows the comparison of LAMOST \texttt{PhDLspec} parameters with GALAH DR3. The consistency in $T_{\rm eff}$ and $\log g$ is similar to the above comparison with APOGEE. While the [Fe/H] shows good agreement for stars with $T_{\rm eff}<6000$~K, there are several stars showing large deviation for stars with $T_{\rm eff}>6000$~K, as the LAMOST \texttt{PhDLspec} gives near solar metallicity but GALAH gives a metallicity lower than $-0.5$. Such a deviation is not observed in Figure~\ref{fig:lamapo2}, although it displays the same star sample as Figure~\ref{fig:lamgalah2}, suggesting those outliers are likely attributed to errors in the GALAH measurements. For a number of elements such as V, Cr, Co, and Ni, the Figure shows an elongated distribution as GALAH abundance exhibits larger dispersion (similar to the case of M67 in Figure~\ref{fig:compare_m67_galah}), even with artificially too high values for stars that LAMOST \texttt{PhDLspec} suggests an ordinary value. For heavy elements Ce and Eu, the \texttt{PhDLspec} abundances exhibit larger dispersion, likely due to larger measurement errors from the LAMOST low-resolution spectra.      

\begin{figure*}[htbp]
		\centering
		\includegraphics[width=1\linewidth]{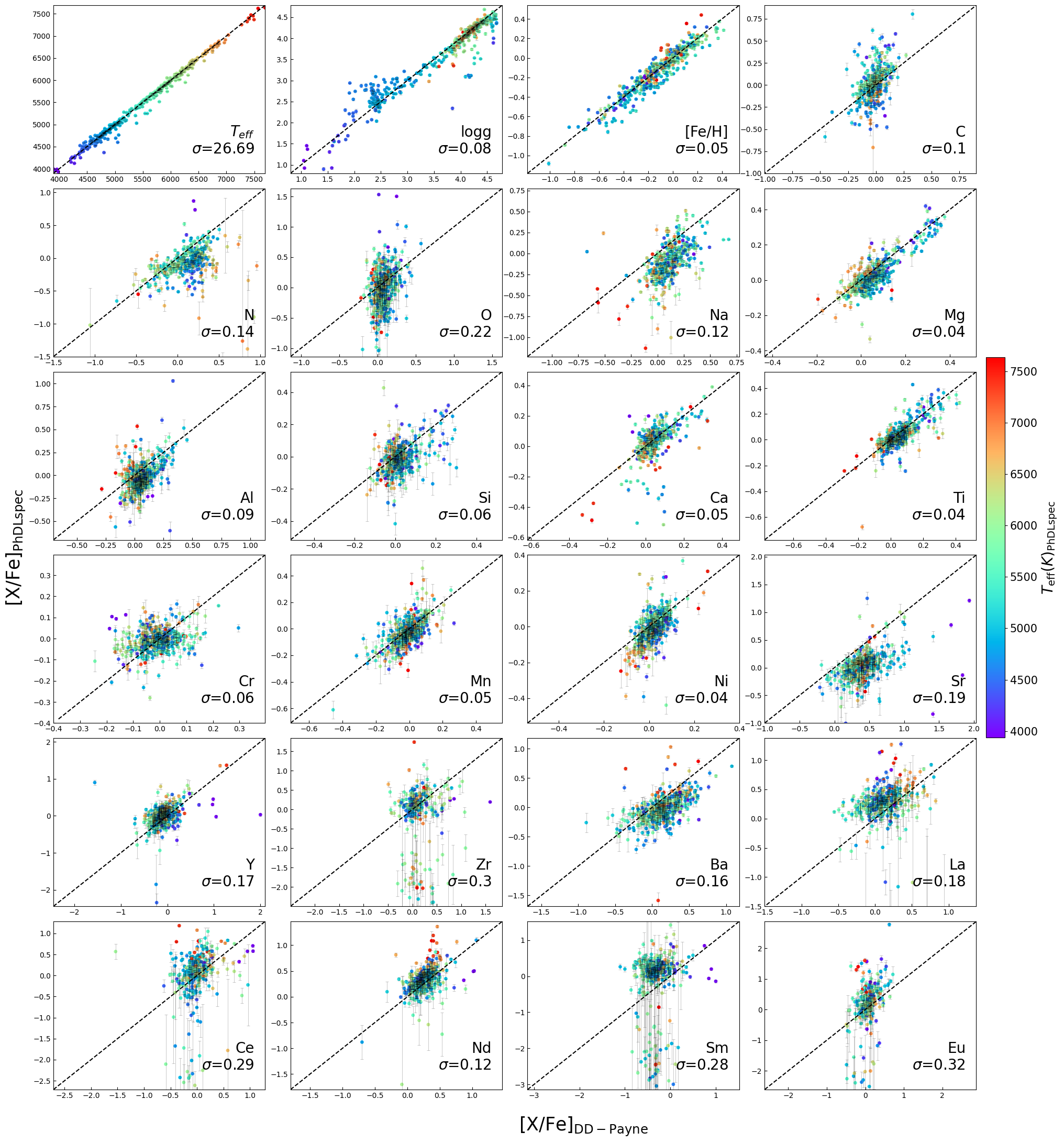}
		\caption{Same as Figure~\ref{fig:lamapo2}, but for comparison with labels derived with \texttt{DD-Payne} from LAMOST spectra by \citet{2025ApJS..279....5Z}.}
       \label{fig:lamddp}
\end{figure*}

We also compared the \texttt{PhDLspec} abundances with the LAMOST \texttt{DD-Payne} estimates for these common star sample. As shown in Figure~\ref{fig:lamddp}, the atmospheric parameters ($T_{\rm eff}$, $\log g$, [Fe/H]) show good agreement with \texttt{DD-Payne} estimates, with dispersion substantially smaller than those presented in Figure~\ref{fig:lamapo2} and Figure~\ref{fig:lamgalah2}. This is also true for most of the elemental abundances. Such a better agreement between measurements based on spectra using different methods than agreement between measurements from different survey datasets may imply that some extra, underestimated systematic errors exist in the survey datasets. Stars with relatively hot temperature ($T_{\rm eff}\gtrsim6000$~K) may show pronounced differences in their abundances between LAMOST \texttt{PhDLspec} and \texttt{DD-Payne} estimates, especially for heavy elements such as Sr, La, Nd, and Eu. This is attributed to the fact that the \texttt{DD-Payne} estimates suffer large errors due to the lack of training set in this regime \citep{2025ApJS..279....5Z}. For some elements such as Sr, Nd, and Sm, the \texttt{PhDLspec} results show a systematic offset to the \texttt{DD-Payne} values, which is a consequence of different zero-point calibration. 

\section{The LAMOST subgiant abundance catalog}\label{sec:subgiant catalog}

\begin{figure*}[htbp]
		\centering
		\includegraphics[width=1\linewidth]{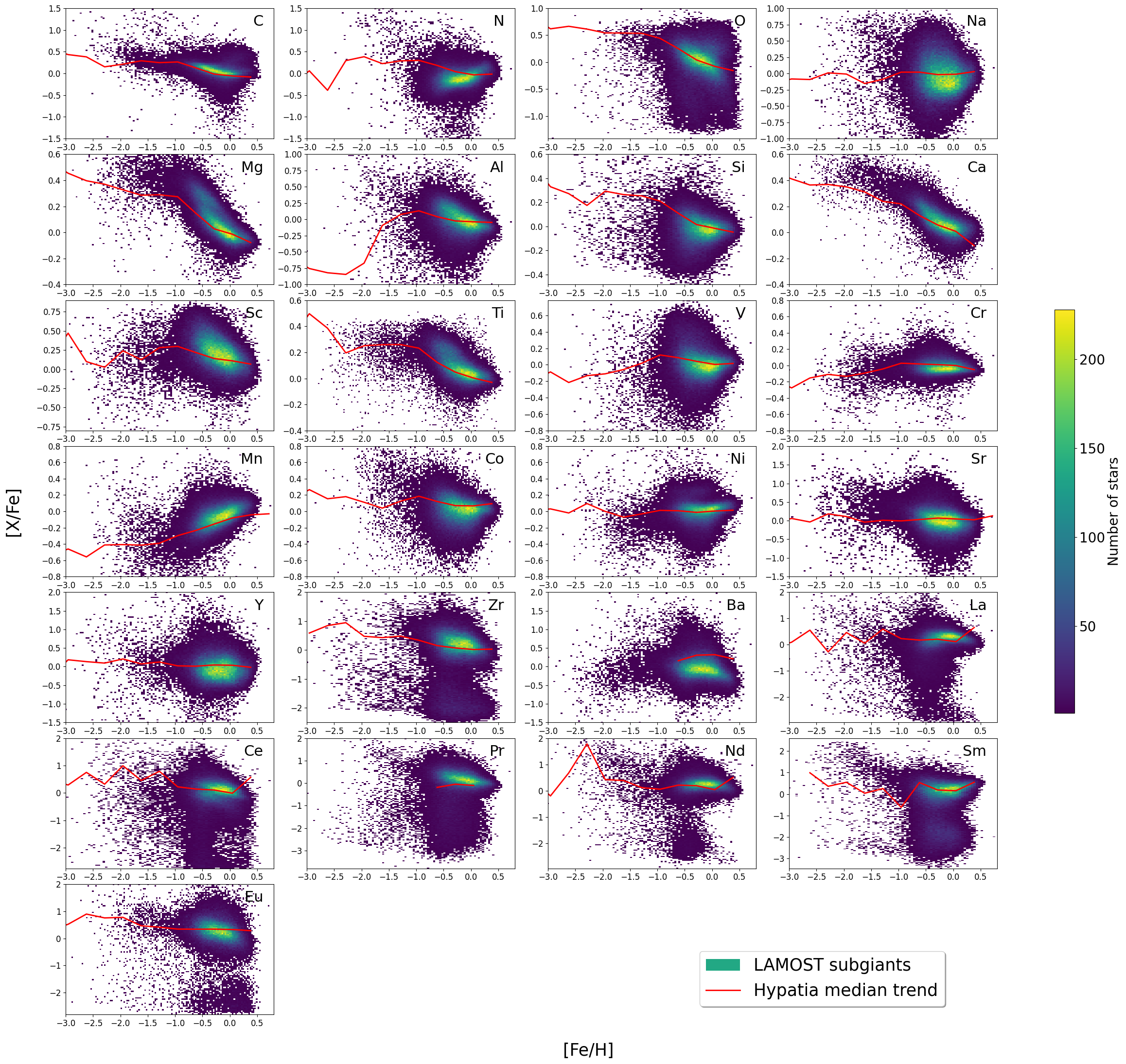}
		\caption{Density scatter plot of elemental abundances derived with \texttt{PhDLspec} from LAMOST spectra for a sample of 116,611 subgiant stars. The red dashed line in each panel indicates the median abundance trend from high-resolution spectroscopy for stars compiled in the Hypatia catalog \citep{2014AJ....148...54H}.}
       \label{fig:subgiant_1}
\end{figure*}
We apply \texttt{PhDLspec} to derive elemental abundances of subgiants in the LAMOST survey. Subgiants represent a transitional evolutionary phase between the main sequence (core hydrogen burning) and the red giant branch (shell hydrogen burning). Their luminosities are highly sensitive to the initial stellar mass -- and thus to main-sequence lifetime and stellar age -- making them ideal targets for Galactic archaeology, as their ages can be determined with high precision. \citet{2022Natur.603..599X} identified 247,104 subgiants using Gaia DR3 and LAMOST DR7 data and estimated their ages with a typical relative precision of 7\%. In this work, we use \texttt{PhDLspec} to derive abundances from LAMOST spectra for 116,611 subgiants with spectral signal-to-noise ratios greater than 50.

Figure~\ref{fig:subgiant_1} shows the scatter density plots of the elemental abundances for our subgiant sample. As a guide to the abundance trends, the red solid line indicates the median abundance trend from high-resolution spectroscopy of 13,384 stars compiled in the Hypatia catalog \citep{2014AJ....148...54H}. For most elements, the trends of the \texttt{PhDLspec} measurements agree well with the high-resolution results. Notably, [Mg/Fe] and [Ti/Fe] exhibit clear bimodal distributions, corresponding to the high-$\alpha$ and low-$\alpha$ populations of the Milky Way. Although Si and Ca are also $\alpha$ elements, their bimodal features are less unambiguous, consistent with previous work \citep[e.g.][]{2025ApJS..279....5Z}, possibly related to different evolutionary paths among these elements.
For the iron-peak elements, [Cr/Fe] and [Ni/Fe] show nearly flat trends with [Fe/H], whereas [Mn/Fe] shows a clear positive trend with [Fe/H]. All trends are qualitatively consistent with high-resolution spectroscopy.  

One notable caveat is the substantial scatter in abundances for many elements—such as N, O, Na, Al, Si, Sc, Ti, V, Y, Zr, La, Ce, Pr, and Eu—among metal-poor stars (${\rm [Fe/H]}\lesssim-1.0$). This scatter arises primarily from large measurement uncertainties, as the spectral features of these elements are too weak in LAMOST's low-resolution spectra to yield robust measurements for metal-poor stars. Furthermore, for a subset of elements (Zr, La, Ce, Pr, Nd, Sm, and Eu), we identify distinct stellar populations exhibiting exceptionally low abundances compared to the bulk sample. The origin of these populations is not yet clear, but we anticipate that they may be artifacts of the measurement process.  

The abundance catalog of the subgiant sample is publicly available via https://doi.org/10.5281/zenodo.19046403. Table~\ref{tab:subgaint catalog} summarizes its contents.
\begin{table}[ht]
\centering
\caption{Description of the subgiant catalog. }
\begin{tabular}{ll}
\toprule
Field & Description \\
\midrule
RA&R.A. from the LAMOST catalog (deg; J2000)\\
DEC&Decl. from the LAMOST catalog (deg; J2000)\\
SOURCEID&Gaia DR3 source ID\\
SPECID&LAMOST spectral ID, in format of \\&obsdate-planid-spid-fiberid\\
UQFLAG&Flag of the repeat visit spectra: 1 means \\&unique star and $>$1 means repeat visits\\ 
$T_{\rm eff}$&Effective temperature (K)\\
$T_{\rm eff}\_\text{err}$&Uncertainty in $T_{\rm eff}$ (K)\\
$\log g$&Surface gravity\\
$\log g\_\text{err}$&Uncertainty in $\log g$\\
$\text{[Fe/H]}$&Iron abundance\\
$\text{[Fe/H]\_err}$&Uncertainty in $\text{[Fe/H]}$\\
$\text{[X/Fe]}$&Element-to-iron abundance after calibration\\
$\text{[X/Fe]\_err}$&Uncertainty in $\text{[X/Fe]}$\\
$\text{[X/Fe]\_flag}$&Flag describing the $\text{[X/Fe]}$ estimation quality\\
$\text{Chisq}$&The $\chi^2$ of the best spectral fitting\\
$\text{[X/Fe]\_uncali}$&Element-to-iron abundance before calibration\\
\bottomrule
\end{tabular}
\label{tab:subgaint catalog}
\end{table}

\section{Summary}\label{sec:summary}
Precise determination of astrophysical parameters from heavily blended spectral features is a key challenge in large-scale spectroscopic surveys with low-to-intermediate spectral resolution. It requires accurate spectral modeling and reliable parameter inference across a high-dimensional label space. 

To address this problem, we have developed \texttt{PhDLspec}, a physical-prior embedded deep learning framework for spectroscopic determination of stellar parameters from spectra. By integrating a transformer-based neural network with physical priors -- specifically the spectral gradients derived from \textit{ab initio} \texttt{Kurucz} atmospheric models -- \texttt{PhDLspec} unites the interpretability of physics-driven modeling with the flexibility and efficiency of modern deep learning. Trained on only a few thousand spectra, \texttt{PhDLspec} can simultaneously model over 30 parameter dimensions to deliver precise predictions of spectra, including both fluxes and their gradients with respect to stellar parameters. With its flexible and precise spectral modeling, \texttt{PhDLspec} simultaneously infers atmospheric parameters ($T_{\rm eff}$, $\log g$, [Fe/H]) and abundances of $\sim30$ elements -- spanning light, $\alpha$-, iron-peak, and neutron-capture groups -- from a single observed spectrum by rigorously utilizing their intrinsic physical features and employing either MCMC or CMA-ES optimization.

The \texttt{PhDLspec} model is lightweight and affordable, enabling training and computation on modest GPU machines, such as the RTX 4090. As a first step toward applying it to the analysis of LAMOST low-resolution ($R<1800$) spectra, we run \texttt{PhDLspec} on calibration star sets, including wide binaries, M67 member stars, and stars in common with the APOGEE and GALAH high-resolution surveys. Validations on these calibration datasets suggest that the parameter estimates achieve high internal precision but suffer substantial systematic errors. After removing systematic trends through a dedicated calibration procedure, the measurement errors, as indicated by dispersion of M67 member stars and wide binaries, are smaller than 0.05~dex in [Fe/H] and abundance ratios [X/Fe] for several elements, including Mg, Si, Ca, Ti, Cr, and Ni; 0.05--0.1~dex in abundance ratios for C, N, Al, Sc, V, Mn, Co, Ba, and Nd; 0.10-0.15~dex for Na, O, Sr, Y, and La; and 0.20-0.30~dex for Zr, Ce, Pr, Sm, Eu.    

We applied \texttt{PhDLspec} to the LAMOST spectra of the subgiant sample from \citet{2022Natur.603..599X}, producing a catalog of 25 elemental abundances for 116,611 stars with precise stellar ages. This catalog is publicly available and is expected to serve as a valuable dataset for unraveling the evolutionary history of the Milky Way.

\texttt{PhDLspec} provides a general and scalable paradigm for physics-grounded spectral modeling and high-dimensional parameter determination. Its capabilities offer critical insights for future spectroscopic analysis, making it particularly crucial for handling forthcoming large spectroscopic datasets. To facilitate these studies, we have made the PhDLspec code publicly available via https://doi.org/10.5281/zenodo.19045278.


\section{Acknowledgements}
We acknowledge financial support from the National Natural Science Foundation of China (NSFC) Grant No. 12588202, the National Key R\&D Program of China Grant No.2022YFF0504200, and NSFC Grant No. 2022000083, 12303025.

This work has used the ATLAS12 stellar atmospheric model from the Kurucz website (http://kurucz.harvard.edu). We acknowledge Robert L. Kurucz and his team for making their model program publicly available and for their continued efforts in maintaining and updating the model.

This work has used spectra data from the Guoshoujing Telescope (LAMOST). LAMOST is a National Major Scientific Project built by the Chinese Academy of Sciences. Funding for the project has been provided by the National Development and Reform Commission. LAMOST is operated and managed by the National Astronomical Observatories, Chinese Academy of Sciences. The LAMOST website is https://www.lamost.org. 


\bibliography{sample631}{}
\bibliographystyle{aasjournal}

\begin{figure*}[htbp]
		\centering
		\includegraphics[width=0.9\linewidth]{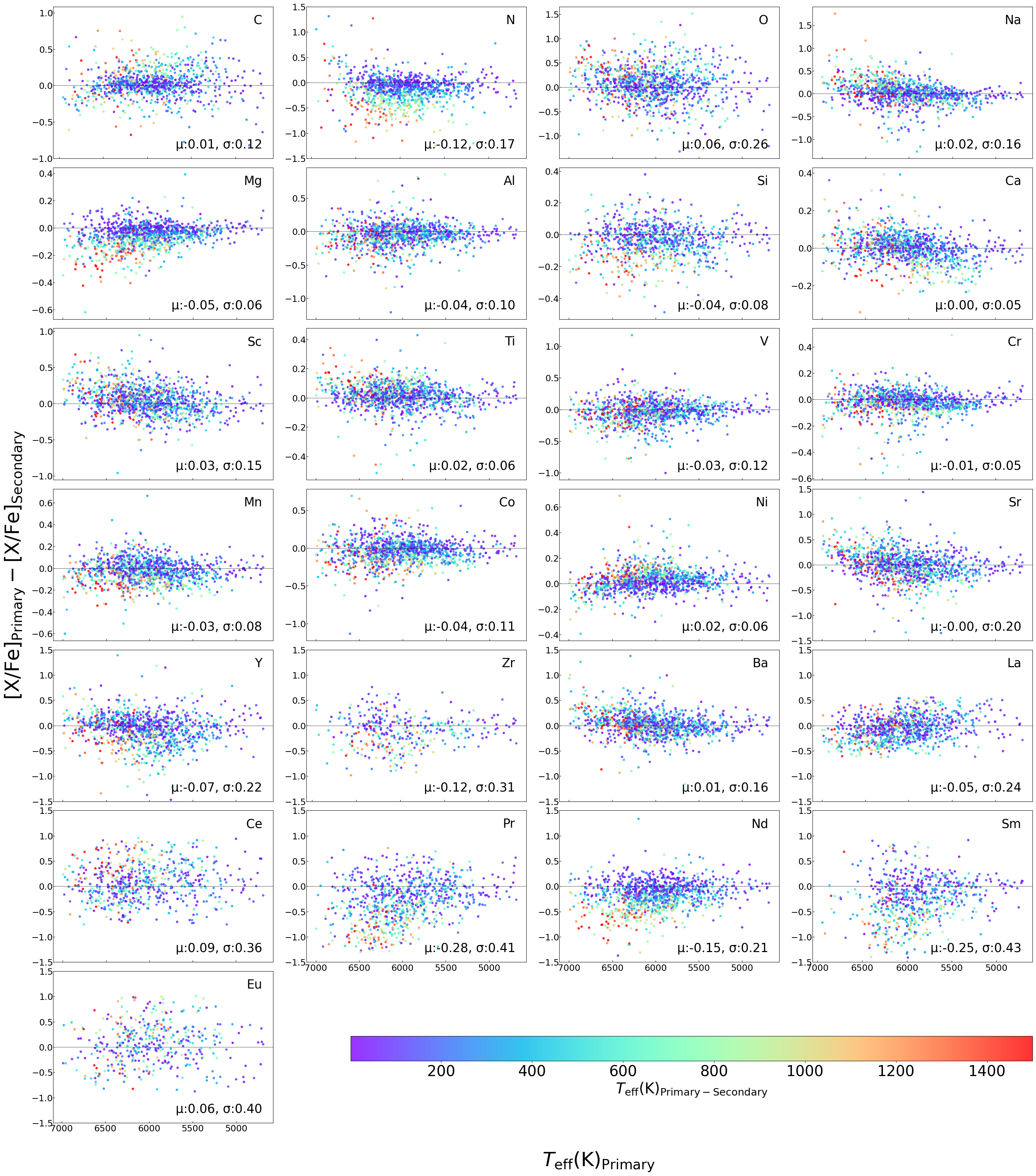}
		\caption{Differences in elemental abundances between the two components of wide binaries, plotted against the effective temperature of the primary star. Each panel shows $\Delta[\mathrm{X/Fe}] = [\mathrm{X/Fe}]_{\rm Primary} - [\mathrm{X/Fe}]_{\rm Secondary}$ for a given element. The color scale indicates the effective temperature difference between the binary components. Systematic trends with $T_{\rm eff}$ are apparent for several elements, particularly Na, Mg, Sr, Pr, and Nd.
}
       \label{fig:widebinary_trend}
\end{figure*}

\appendix
While Figure~\ref{fig:widebinary} in the main text presents the difference in the calibrated elemental abundances between binary component stars, we show here in Figure~\ref{fig:widebinary_trend} the corresponding results for the original \texttt{PhDLspec} estimates--i.e., abundances prior to calibration. For several elements, such as N, Mg, Si, Zr, Pr, and Nd, visible trends emerge as a function of the primary star's temperature or the temperature difference between the components. These trends underscore the necessity of applying the calibration developed in this work.
\end{document}